\documentclass[reprint,superscriptaddress,amsmath,amssymb,aps,prx,floatfix,longbibliography]{revtex4-2}

\usepackage{graphicx}
\usepackage{amsmath,amssymb}
\usepackage[dvipsnames]{xcolor}
\usepackage{hyperref}
\hypersetup{hidelinks}
\usepackage{placeins}

\newcommand{\Appref}[1]{Appendix~\ref{#1}}
\newcommand{\maybeincludegraphics}[2][0.95\linewidth]{%
\IfFileExists{#2}{\includegraphics[width=#1]{#2}}{%
\begin{center}
\fbox{\begin{minipage}{0.86\linewidth}
\centering\vspace{0.8cm}
\textbf{Missing figure file: \texttt{\detokenize{#2}}}\\[4pt]
Insert the final figure here, or replace the file name in the source.\vspace{0.8cm}
\end{minipage}}
\end{center}}}

\begin{document}

\title{Floquet-Resolved Dissipation Selects Entanglement Beyond Population Spectroscopy}

\author{Gianluigi Pacino}
\affiliation{Dipartimento di Fisica e Astronomia, Universit{\`a} di Catania, 95123 Catania, Italy}
\affiliation{Centro Siciliano di Fisica Nucleare e Struttura della Materia, Catania, Italy}
\author{Rosario Nicosia}
\affiliation{Dipartimento di Fisica e Astronomia, Universit{\`a} di Catania, 95123 Catania, Italy}
\affiliation{Centro Siciliano di Fisica Nucleare e Struttura della Materia, Catania, Italy}
\author{Alessandro Ridolfo}
\email{Contact author: alessandro.ridolfo@dfa.unict.it}
\affiliation{Dipartimento di Fisica e Astronomia, Universit{\`a} di Catania, 95123 Catania, Italy}
\affiliation{INFN, Sez. Catania, 95123, Catania, Italy}

\begin{abstract}
Reliable quantum-state engineering in periodically driven devices requires more than reproducing their excitation spectrum: the environment must resolve the transitions of the driven system. We show that two dissipative descriptions of the same parametrically coupled qubits can yield closely similar period-averaged populations yet qualitatively different asymptotic entanglement. Both retain the complete periodically driven Hamiltonian, the same microscopic bath couplings, and the same physical observables; they differ in the dynamical representation used to resolve the dissipative channels and the corresponding system operators: the static dressed basis in the partial harmonic decomposition approach (PHDA) and the Floquet basis, including all relevant drive sidebands, in Floquet-Born-Markov (FBM) theory. The population maps preserve the same resonance skeleton and remain closely similar over broad low-to-moderate drive regions, while clearer differences emerge as the modulation becomes stronger. Phase-resolved single- and two-qubit coherence observables reveal a substantially stronger redistribution of amplitudes and phases. Concurrence amplifies this hidden state-level discrepancy: Floquet-resolved dissipation selects stronger and more extended Bell-like entangled regions and a larger overlap with a fixed Bell-like reference, while PHDA generally underestimates the entanglement and its thermal persistence. A channel-resolution test verifies the secular, completely positive FBM construction throughout the relevant parameter domain. Our results establish population agreement as an insufficient benchmark for open Floquet quantum-state engineering and identify coherence-sensitive observables as the decisive validation test.
\end{abstract}

\date{\today}
\maketitle

\section{Introduction}

Periodic modulation is a central tool for activating otherwise off-resonant interactions, implementing tunable two-qubit gates, and generating entanglement in superconducting and spin-qubit platforms~\cite{Reagor2018,Hong2020,Sung2021,McKay2016,Roth2017,Caldwell2018,Ganzhorn2020,Sete2021,Song2020,Gallardo2022,Srinivasa2024}. Its closed-system design is encoded in the quasienergy spectrum and micromotion of the driven Hamiltonian~\cite{Shirley1965,Sambe1973,Grifoni1998,Bukov2015,Silveri2017,Eckardt2017,Oka2019,Mori2023}. In an open device, however, the same modulation also reorganizes the transition frequencies and matrix elements sampled by the environment. Here we show that neglecting this reorganization can remain almost invisible in population spectroscopy while producing a qualitatively different entanglement prediction.

This failure mode matters because dissipation does not merely attenuate a Floquet-engineered state: it selects the asymptotic periodic state from the available drive-dressed pathways. Population maps predominantly reveal the common coherent resonance skeleton and can therefore agree even when two generators resolve different transition spectra. Phase-sensitive coherences retain the relative amplitudes and phases of those pathways, while entanglement is more demanding still because concurrence compares off-diagonal matrix elements with population-dependent separability thresholds. Agreement at the level of occupations can consequently coexist with substantial differences in the coherences and, hence, in coherence-sensitive quantities such as entanglement. This perspective connects Floquet engineering with the broader use of reservoirs to prepare and autonomously stabilize nonclassical states~\cite{Poyatos1996,Kraus2008,Verstraete2009,Krauter2011,Lin2013,Shankar2013,Reiter2013,Harrington2022}.

We isolate this mechanism in two transversely coupled qubits whose interaction is modulated as $A\cos(\omega_d t)\sigma_x^{(1)}\sigma_x^{(2)}$. The modulation hybridizes excitation sectors at avoided quasienergy crossings and makes Bell-like superpositions dynamically accessible. We compare two constructions with the same system Hamiltonian $H_S(t)$, the same local system-bath coupling operators, and the same bosonic reservoirs. In the partial harmonic decomposition approach (PHDA), the full coherent time dependence is retained and the periodic density matrix is solved in drive harmonics, while the dissipative transitions and corresponding dressed system operators are represented in the eigenbasis of the static interacting Hamiltonian $H_0$. In Floquet-Born-Markov (FBM) theory, the same physical coupling and system operators are instead resolved between Floquet modes of the complete driven Hamiltonian, including all relevant sidebands. In the low-excitation regime, where the drive-induced dressing of the dissipative channels remains weak, the PHDA provides an excellent approximation to the FBM treatment. It should therefore be regarded as an efficient approximate construction of the dissipative dynamics within this regime, rather than as a different model of the coherently driven system. Static dressed-state and Floquet-resolved treatments of dissipation and photodetection provide the respective theoretical foundations~\cite{Beaudoin2011,Ridolfo2012,Garziano2013,Garziano2017,Settineri2018,Macri2022,Mercurio2022,FornDiaz2019,Kockum2019,Kohler1997,Grifoni1998,Gasparinetti2013,Hone2009,Mori2023}.

The comparison reveals a pronounced hierarchy. Period-averaged dressed populations remain closely similar over broad low-to-moderate drive regions, while clearer differences appear once the modulation becomes sufficiently strong. At a fixed drive phase, the single-qubit and mixed coherence-sensitive observables retain the same sequence of sign-changing resonances, while displaying substantially different amplitudes, widths, and relative weights. Concurrence converts these hidden coherence changes into a qualitative state-level distinction: FBM selects stronger and more extended Bell-like entangled domains that PHDA generally underestimates. The same Floquet-resolved states display a different and generally larger overlap with a fixed Bell-like reference, and the concurrence advantage persists under thermal activation.
Recent work on ultrastrong cavity QED has independently established that driven dissipation is governed by quasienergy-resolved channels and has developed a fully nonsecular Floquet generalized master equation for populations, spectra, and Floquet-Liouville modes~\cite{Akbari2026}. Our work addresses the complementary question of stationary two-qubit coherence and entanglement in the weak-damping, spectrally resolved regime. Secularization in the total Floquet transition frequency produces a Gorini-Kossakowski-Lindblad-Sudarshan (GKLS) generator and hence completely positive, trace-preserving dynamics~\cite{Gorini1976,Lindblad1976}. A direct channel-resolution test shows that the broad coherence and entanglement domains lie safely within this regime, with unresolved channels confined to narrow near-degeneracy loci. Therefore, the resulting message is general: reproducing driven occupations is not sufficient to validate an open-system approximation intended for quantum-state engineering.

\begin{figure*}[!t]
\maybeincludegraphics[0.8\textwidth]{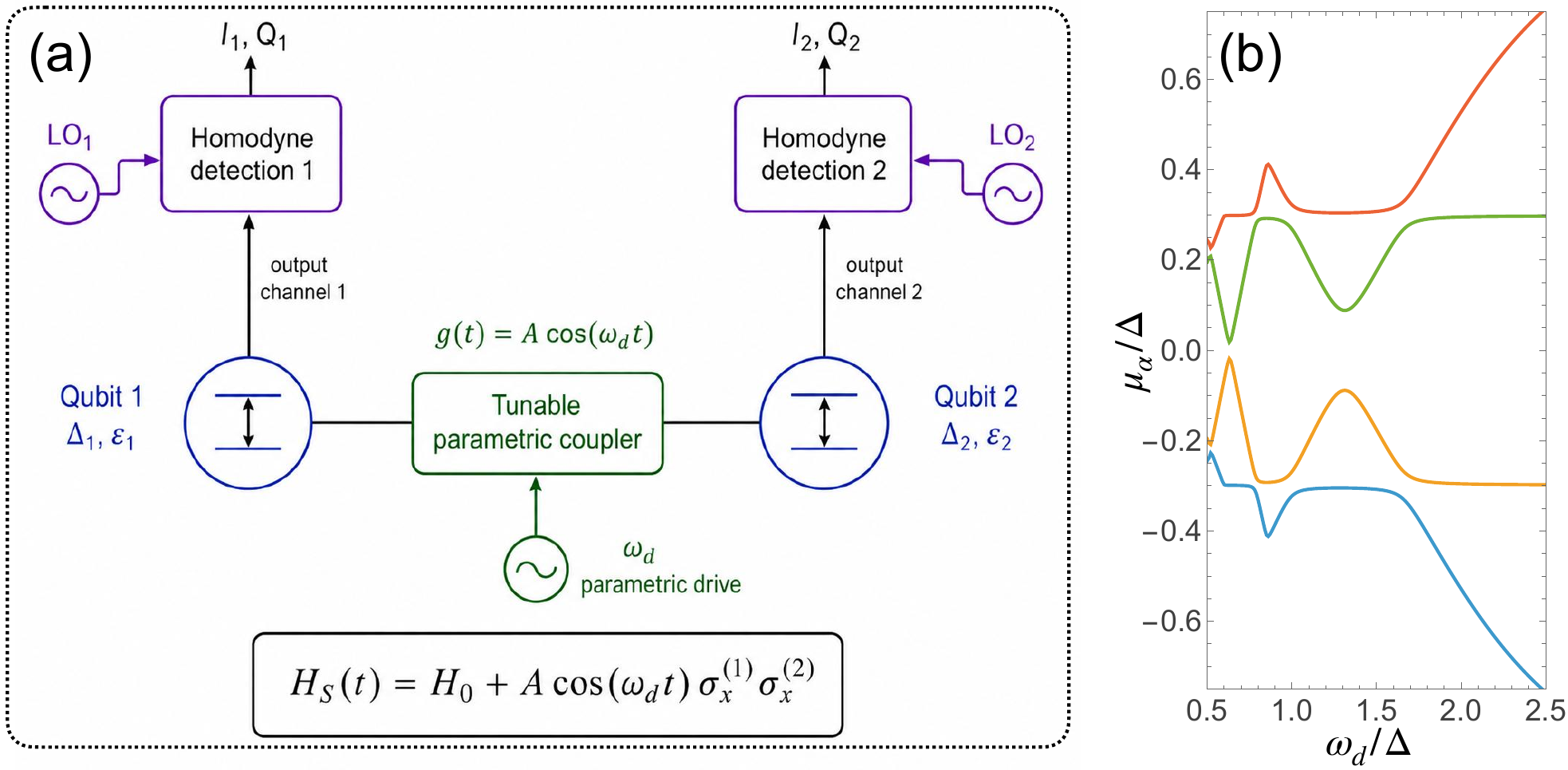}
\caption{\textbf{Parametrically coupled qubits and Floquet spectrum.} (a) Minimal scheme of two effective qubits with static transverse coupling contained in $H_0$ and a tunable interaction modulated as $g(t)=A\cos(\omega_d t)$. Each qubit is coupled to an independent output channel, shown with phase-sensitive homodyne detection of the two quadratures. (b) Four Floquet quasienergies $\mu_\alpha$, folded into the first Brillouin zone, as a function of $\omega_d/\Delta$. Avoided crossings mark the multiphoton resonances at which the drive hybridizes different excitation sectors. Parameters are $\Delta_1=\Delta_2=\Delta$, $\epsilon_1/\Delta=0.1$, $\epsilon_2/\Delta=0$, $g_0/\Delta=0.3$, and $A/\Delta=1$.}
\label{fig:setup_quasienergies}
\end{figure*}

\section{Model and Floquet-resolved dissipation}
\label{sec:model}

To isolate the role of dissipative channel resolution, both descriptions use the same system Hamiltonian, reservoirs, and microscopic coupling operators. In what follows we set $\hbar=k_{\rm B}=1$. The system consists of two effective two-level emitters with a static transverse coupling and a parametric modulation of the same interaction channel,
\begin{align}
H_S(t)&=H_0+A\cos(\omega_d t)\,\sigma_x^{(1)}\sigma_x^{(2)},\notag\\
H_0&=\frac{1}{2}\sum_{i=1}^{2}\left(\Delta_i\sigma_z^{(i)}+\epsilon_i\sigma_x^{(i)}\right)
+g_0\sigma_x^{(1)}\sigma_x^{(2)} .
\label{eq:system_hamiltonian}
\end{align}
Here, $i=1,2$ labels the two qubits, $\sigma_{\alpha}^{(i)}$ ($\alpha=x,z$) are the corresponding Pauli operators, and $\Delta_i$ and $\epsilon_i$ denote the longitudinal and transverse energy scales entering the local qubit Hamiltonians, respectively. The parameter $g_0$ is the static transverse coupling strength, while $A$ and $\omega_d$ are, respectively, the amplitude and angular frequency of the periodic modulation of the interqubit coupling. Accordingly, $H_0$ denotes the complete undriven Hamiltonian, including both the local qubit terms and their static interaction.
The two qubits couple locally to independent bosonic reservoirs. Before any decomposition into upward and downward transitions, the common microscopic Hamiltonian is
\begin{align}
H_{\rm tot}(t)&=H_S(t)+H_B+H_{SB},\notag\\
H_B&=\sum_n\omega_n^{(1)}b_n^\dagger b_n
+\sum_n\omega_n^{(2)}c_n^\dagger c_n,\notag\\
H_{SB}&=\sum_n g_n^{(1)}(b_n+b_n^\dagger)\sigma_x^{(1)}
+\sum_n g_n^{(2)}(c_n+c_n^\dagger)\sigma_x^{(2)},
\label{eq:total_hamiltonian}
\end{align}
where $H_{\rm tot}(t)$ denotes the total Hamiltonian of the system and its environment, with $H_B$ describing the two independent bosonic reservoirs and $H_{SB}$ their interaction with the qubits. The operators $b_n$ ($b_n^\dagger$) and $c_n$ ($c_n^\dagger$) annihilate (create) an excitation in the $n$th mode of the reservoirs coupled to qubits 1 and 2, respectively, while $\omega_n^{(i)}$ denotes the corresponding bath-mode frequency and $g_n^{(i)}$ the microscopic system--bath coupling strength. Each reservoir couples locally to its qubit through the physical transverse operator $\sigma_x^{(i)}$.
Equation~\eqref{eq:total_hamiltonian} does not yet privilege either approximation: in both calculations the reservoirs couple to the same physical operators $\sigma_x^{(i)}$. The weak coupling between each qubit and its reservoir permits the same system--bath RWA in both cases, and the distinction is the basis used to define the raising and lowering components.

For PHDA, let $U_0$ be the unitary matrix whose columns are the eigenstates of the undriven interacting Hamiltonian $H_0$, so that $U_0^\dagger H_0 U_0$ is diagonal. In the following, operator matrices expressed in this static dressed representation will be denoted by the subscript $d$. The frequency-resolved transition operators are defined by
\begin{align}
H_{0,\rm d}&\equiv U_0^\dagger H_0U_0=\sum_jE_j^{(0)}|j\rangle\langle j|,\notag\\
\sigma_{x,\rm d}^{(i)}&\equiv U_0^\dagger\sigma_x^{(i)}U_0,\notag\\
S_{i,\nu}^{-}&=\sum_{\substack{j,\ell\\E_\ell^{(0)}-E_j^{(0)}=\nu}}
\Theta\!\left(E_\ell^{(0)}-E_j^{(0)}\right)
\langle j|\sigma_{x,\rm d}^{(i)}|\ell\rangle|j\rangle\langle\ell|,\notag\\
S_{i,\nu}^{+}&=[S_{i,\nu}^{-}]^\dagger,\qquad
S_i^\pm=\sum_{\nu>0}S_{i,\nu}^\pm ,
\label{eq:static_dressed_operators}
\end{align}
where $|j\rangle$ and $|\ell\rangle$ denote the basis states associated with the eigenstates of the undriven interacting Hamiltonian $H_0$ in the static dressed representation, with corresponding eigenenergies $E_j^{(0)}$ and $E_\ell^{(0)}$. We take $\Theta(x)=1$ for $x>0$ and $\Theta(x)=0$ for $x\leq0$, so that diagonal matrix elements with exactly zero transition energy are excluded. 
By construction, $S_{i,\nu}^{-}$ and $S_{i,\nu}^{+}$ represent, respectively, the dressed lowering and raising components associated with transitions of frequency $\nu$: $S_{i,\nu}^{-}$ maps an upper dressed state onto a lower one, whereas $S_{i,\nu}^{+}=(S_{i,\nu}^{-})^\dagger$ generates the reverse transition. The operators $S_i^{-}=\sum_{\nu}S_{i,\nu}^{-}$ and $S_i^{+}=(S_i^{-})^\dagger=\sum_{\nu}S_{i,\nu}^{+}$ are the corresponding full dressed lowering and raising parts of the physical operator $\sigma_x^{(i)}$, collecting all allowed downward and upward transitions of qubit $i$, respectively.

The coherent modulation is nevertheless retained nonperturbatively: the periodic steady state is obtained by expanding the full time-dependent master equation in drive harmonics and solving the resulting time-independent block problem. PHDA therefore freezes only the transition decomposition used by dissipation, not the driven evolution. In FBM, by contrast, the same microscopic operators are decomposed only after solving the complete Floquet problem,
\begin{align}
\mathrm{PHDA}:\quad &\sigma_{x,\rm d}^{(i)}
\xrightarrow{\;H_0\text{ eigenbasis}\;}\{S_{i,\nu}^{\pm}\},\notag\\
\mathrm{FBM}:\quad &[H_S(t)-i\partial_t]|\Phi_\alpha(t)\rangle
=\mu_\alpha|\Phi_\alpha(t)\rangle,\notag\\[-2pt]
&\hspace{1.2cm}\sigma_x^{(i)}
\xrightarrow{\;\text{Floquet basis}\;}\{\widetilde S_{i,k}^{\pm}(\omega)\}.
\label{eq:phda_fbm_comparison}
\end{align}
Here $|\Phi_\alpha(t)\rangle$ are $\tau$-periodic Floquet modes and $\mu_\alpha$ are their quasienergies, defined modulo $\omega_d$.
Figure~\ref{fig:setup_quasienergies}(a) summarizes the minimal architecture. It is not tied to a unique hardware realization: parametrically driven tunable couplers and modulated exchange elements provide natural circuit-QED implementations~\cite{Reagor2018,Hong2020,Sung2021,McKay2016,Roth2017,Caldwell2018,Ganzhorn2020,Sete2021}. The output channels indicate how the system observables considered below can be accessed experimentally. The main-text maps show dressed populations and coherence-sensitive system observables; calibrated emission rates or homodyne signals additionally contain the spectral weighting of the monitored line. The corresponding input-output relation and the PHDA and FBM representations used to evaluate its system source term are given in \Appref{app:floquet}, where the reservoir spectrum and an independent detector response are kept distinct. Figure~\ref{fig:setup_quasienergies}(b) shows the avoided quasienergy crossings generated by the modulation. They form the common coherent resonance skeleton of both calculations, so the differences reported below originate from how the environment resolves that skeleton and, for output observables, from the corresponding dressed representation of the detection operators.

Untilded $S_{i,\nu}^{\pm}$ denote the static dressed operators used in PHDA, whereas tilded quantities are constructed from the full Floquet propagator. For both PHDA and FBM we use an operator-action convention: $+$ raises the corresponding dressed transition and $-$ lowers it. The harmonic-space PHDA construction, the Floquet interaction-picture decomposition, and their use in the input-output source operators are detailed in \Appref{app:floquet}.
For the FBM construction we evolve the microscopic coupling operators $\sigma_x^{(i)}$ with the full driven propagator $U_S(t)$, defined by
\begin{equation}
i\partial_t U_S(t)=H_S(t)U_S(t),
\qquad U_S(0)=\mathbb I ,
\label{eq:system_propagator}
\end{equation}
whose formal solution is the time-ordered exponential
$U_S(t)=\mathcal{T}\exp[-i\int_0^t H_S(t')\,dt']$, and can be obtained numerically by integrating the time-dependent Schrödinger equation~\cite{Bluemel1991}. The Floquet modes satisfy
$U_S(t)|\Phi_\alpha(0)\rangle=e^{-i\mu_\alpha t}|\Phi_\alpha(t)\rangle$, with $|\Phi_\alpha(t+\tau)\rangle=|\Phi_\alpha(t)\rangle$ and $\tau=2\pi/\omega_d$. 
In terms of a complete set of Floquet modes, the same propagator can be written as $U_S(t)=\sum_\alpha e^{-i\mu_\alpha t}
|\Phi_\alpha(t)\rangle\langle\Phi_\alpha(0)|$.
For convenience, the Floquet quasienergies $\mu_\alpha$ are folded here into the first Brillouin zone $[-\omega_d/2,\omega_d/2]$. The exact interaction-picture operator is
\begin{align}
\widetilde\sigma_x^{(i)}(t)
&=U_S^\dagger(t)\sigma_x^{(i)}U_S(t)\notag\\
&=\sum_{\alpha\beta k} e^{i(\mu_\alpha-\mu_\beta+k \, \omega_d)t}
X_{\alpha\beta}^{(i)}(k)|\Phi_\alpha(0)\rangle\langle\Phi_\beta(0)|.
\label{eq:floquet_operator_decomp}
\end{align}
where
\begin{equation}
X_{\alpha\beta}^{(i)}(k)=\frac{1}{\tau}\int_0^\tau dt\,
\langle\Phi_\alpha(t)|\sigma_x^{(i)}|\Phi_\beta(t)\rangle e^{-ik \, \omega_d t}.
\end{equation}
Writing $\omega=\mu_\alpha-\mu_\beta$, we collect matrix elements with the same quasienergy difference into
\begin{equation}
X_k^{(i)}(\omega)=
\sum_{\mu_\alpha-\mu_\beta=\omega}
X_{\alpha\beta}^{(i)}(k)|\Phi_\alpha(0)\rangle\langle\Phi_\beta(0)|,
\end{equation}
and define the total transition frequency $\Delta_k(\omega)=\omega+k \, \omega_d$. The Floquet raising and lowering components are
\begin{equation}
\widetilde S_{i,k}^{+}(\omega)=\Theta[\Delta_k(\omega)]X_k^{(i)}(\omega),
\qquad
\widetilde S_{i,k}^{-}(\omega)=[\widetilde S_{i,k}^{+}(\omega)]^\dagger .
\end{equation}
The same convention $\Theta(0)=0$ is used here. Hence an exactly vanishing total transition frequency is not assigned to either the raising or lowering sector; with the Brillouin-zone convention adopted above, this in particular removes the $k=0$, $\omega=0$ channel retained neither in the dissipator nor in the detection decomposition.
Their sums reconstruct the corresponding transition components of the original interaction-picture operator, $\widetilde S_i^\pm(t)=\sum_{\omega k}e^{\pm i\Delta_k(\omega)t}\widetilde S_{i,k}^{\pm}(\omega)$. Under the same weak qubit--reservoir coupling assumption, the system--bath RWA pairs bath annihilation operators with the system-raising components and bath creation operators with the system-lowering components, i.e. the counter-rotating combinations are neglected. We keep the $(\omega,k)$ decomposition because the later secular approximation acts on the individual total transition frequencies $\Delta_k(\omega)$, as the channel labels do not represent a different microscopic system-bath coupling. Details are given in \Appref{app:floquet}.

After the common system-bath RWA, tracing out the independent bosonic reservoirs in the Born-Markov approximation and secularizing between distinct $(\omega,k)$ Floquet channels yields the FBM generator
\begin{widetext}
\begin{equation}
\dot{\tilde\rho}=\mathcal L_{\rm FBM}\tilde\rho
=\sum_{i,\omega,k}\frac{\gamma^{(i)}[\Delta_k(\omega)]}{2}\{1+n[\Delta_k(\omega),T_i]\}\mathcal D[\widetilde S_{i,k}^{-}(\omega)]\tilde\rho
+\sum_{i,\omega,k}\frac{\gamma^{(i)}[\Delta_k(\omega)]}{2}n[\Delta_k(\omega),T_i]\mathcal D[\widetilde S_{i,k}^{+}(\omega)]\tilde\rho .
\label{eq:fbm_master}
\end{equation}
\end{widetext}
where $\mathcal D[O]\rho=2O\rho O^\dagger-\{O^\dagger O,\rho\}$ and $n(\nu,T_i)=[\exp(\nu/T_i)-1]^{-1}$. For an Ohmic bath, $\gamma^{(i)}(\nu)=\gamma_0^{(i)}\times(\nu/\Delta)\times\Theta(\nu)$. For the symmetric reservoirs used throughout the numerical results, we set $\gamma_0^{(1)}=\gamma_0^{(2)}\equiv\gamma$. The rates are evaluated only for channels with positive total transition frequency selected by the Floquet decomposition. The FBM stationary state in the Floquet interaction picture is obtained from
\begin{equation}
\mathcal L_{\rm FBM}\tilde\rho_{\rm ss}=0,\qquad {\rm Tr}\,\tilde\rho_{\rm ss}=1,
\label{eq:stationary_state}
\end{equation}
and it gives the periodic steady state (PSS) $\rho_{\rm pss}^{(\mathrm{FBM})}(t)=U_S(t)\tilde\rho_{\rm ss}U_S^\dagger(t)$. In PHDA, as we work in the Schr\"odinger picture, the PSS is written directly in the fixed static-dressed working representation as
$\rho_{\rm pss}^{(\mathrm{PHDA})}(t)=\sum_n\rho_n e^{in\omega_dt}$, where the matrices $\rho_n$ are the Fourier components obtained from the coupled harmonic equations. Restoring these Fourier phases supplies the entire stroboscopic time dependence; details are given in \Appref{app:floquet}. Although both calculations contain the same time-dependent Hamiltonian $H_S(t)$, they employ different spectral constructions for the dissipative dynamics and, correspondingly, different decompositions of the system and detection operators. 
The FBM approach resolves the dissipative channels in the Floquet basis, including quasienergy sidebands, whereas the PHDA constructs them from $H_0$ and incorporates the drive through the density-matrix harmonics. The dominant resonance structure can therefore remain similar while the static dressed dissipator is accurate, even though the underlying asymptotic states differ. These differences are more clearly exposed by coherence-sensitive observables, for which the FBM captures the drive-induced redistribution of each transition among Floquet sidebands.

\begin{figure*}[!t]
\maybeincludegraphics[0.8\textwidth]{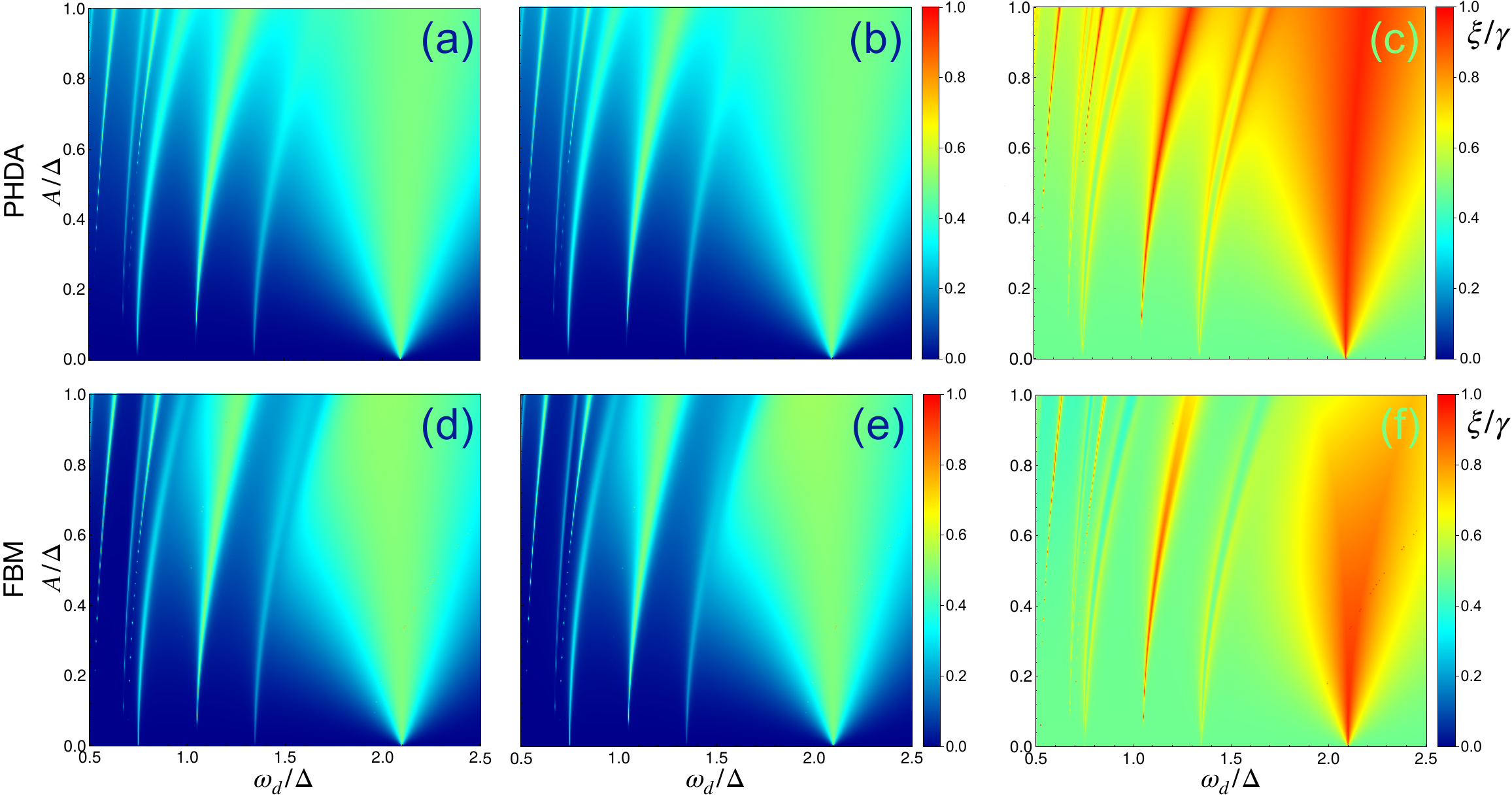}
\caption{\textbf{Period-averaged dressed populations and Liouvillian gap.} The upper (lower) row corresponds to PHDA (FBM). Panels (a,d) and (b,e) show the period-averaged dressed populations of qubits 1 and 2, respectively, while panels (c,f) show the scaled Liouvillian gap $\xi/\gamma$, with $\xi=\min_{j\neq0}[-\mathrm{Re}(\lambda_j)]$. All color scales range from 0 to 1. Parameters are $\Delta_1=\Delta_2=\Delta$, $\epsilon_1/\Delta=0.1$, $\epsilon_2/\Delta=0$, $g_0/\Delta=0.3$, $T_1=T_2=0$, and $\gamma/\Delta=10^{-3}$ for Ohmic reservoirs.}
\label{fig:populations_gap}
\end{figure*}

The channel secularization is a structural element of the present approach, as transitions with the same quasienergy difference at fixed $k$ are grouped in a common jump operator, while distinct $(\omega,k)$ blocks are separated. Since every rate in Eq.~\eqref{eq:fbm_master} is non-negative on the retained channels with positive total transition frequency, $\mathcal L_{\rm FBM}$ is of GKLS form~\cite{Gorini1976,Lindblad1976}, thus $e^{t\mathcal L_{\rm FBM}}$ is completely positive and trace preserving. This distinguishes the present FBM generator from fully nonsecular Floquet-Redfield equations, which retain interchannel interference but are not guaranteed to generate completely positive dynamics~\cite{Akbari2026,Farina2019,NathanRudner2020,Mozgunov2020,Trushechkin2021}. The price of the GKLS construction is the usual secular resolution condition: distinct total Floquet transition frequencies must be separated on the scale of the dissipative linewidths~\cite{Davies1974,Hone2009}. We test this condition directly through the minimum channel-resolution ratio $\mathcal R_{\min}$ defined in \Appref{app:secularity}. For the parameter set analyzed below, $\Delta_1=\Delta_2=\Delta$, $\epsilon_1/\Delta=0.1$, $\epsilon_2=0$, $g_0/\Delta=0.3$, $T_1=T_2=0$, and $\gamma/\Delta=10^{-3}$, with Ohmic reservoirs, the map of $\log_{10}(\mathcal R_{\min})$ is non-negative over essentially the whole $(\omega_d,A)$ plane. Values $\mathcal R_{\min}<1$ occur only on extremely narrow degeneracy loci, which are clipped in the appendix plot. The fully secular approximation is therefore quantitatively well controlled throughout the parameter regions supporting the population, coherence, and entanglement results reported below, and a partial-secular or nonsecular extension would be needed only in the immediate vicinity of the unresolved loci.

Although the numerical results are obtained from the full stationary density matrix in Eq.~\eqref{eq:stationary_state}, a Pauli-type population equation is useful for interpreting the dissipative pathways in the fully secular limit. When populations and coherences decouple, the Floquet-state populations satisfy
\begin{equation}
\dot p_\alpha=\sum_\beta W_{\alpha\leftarrow\beta}p_\beta-
\left(\sum_\gamma W_{\gamma\leftarrow\alpha}\right)p_\alpha,
\label{eq:pauli_population}
\end{equation}
or $\dot{\mathbf p}=M\mathbf p$, with $M_{\alpha\beta}=W_{\alpha\leftarrow\beta}$ for $\alpha\ne\beta$ and $M_{\alpha\alpha}=-\sum_\gamma W_{\gamma\leftarrow\alpha}$. In this reduced picture the stationary populations follow from $M\mathbf p^{\rm ss}=0$ and $\sum_\alpha p_\alpha^{\rm ss}=1$. The channel-resolved rates $W_{\alpha\leftarrow\beta}^{(i,\pm)}(\omega,k)$, their expression in terms of the Fourier amplitudes $X_{\alpha\beta}^{(i)}(k)$, and the limitations of the rate-only interpretation are given in \Appref{app:rates}.

\begin{figure*}[!t]
\maybeincludegraphics[0.8\textwidth]{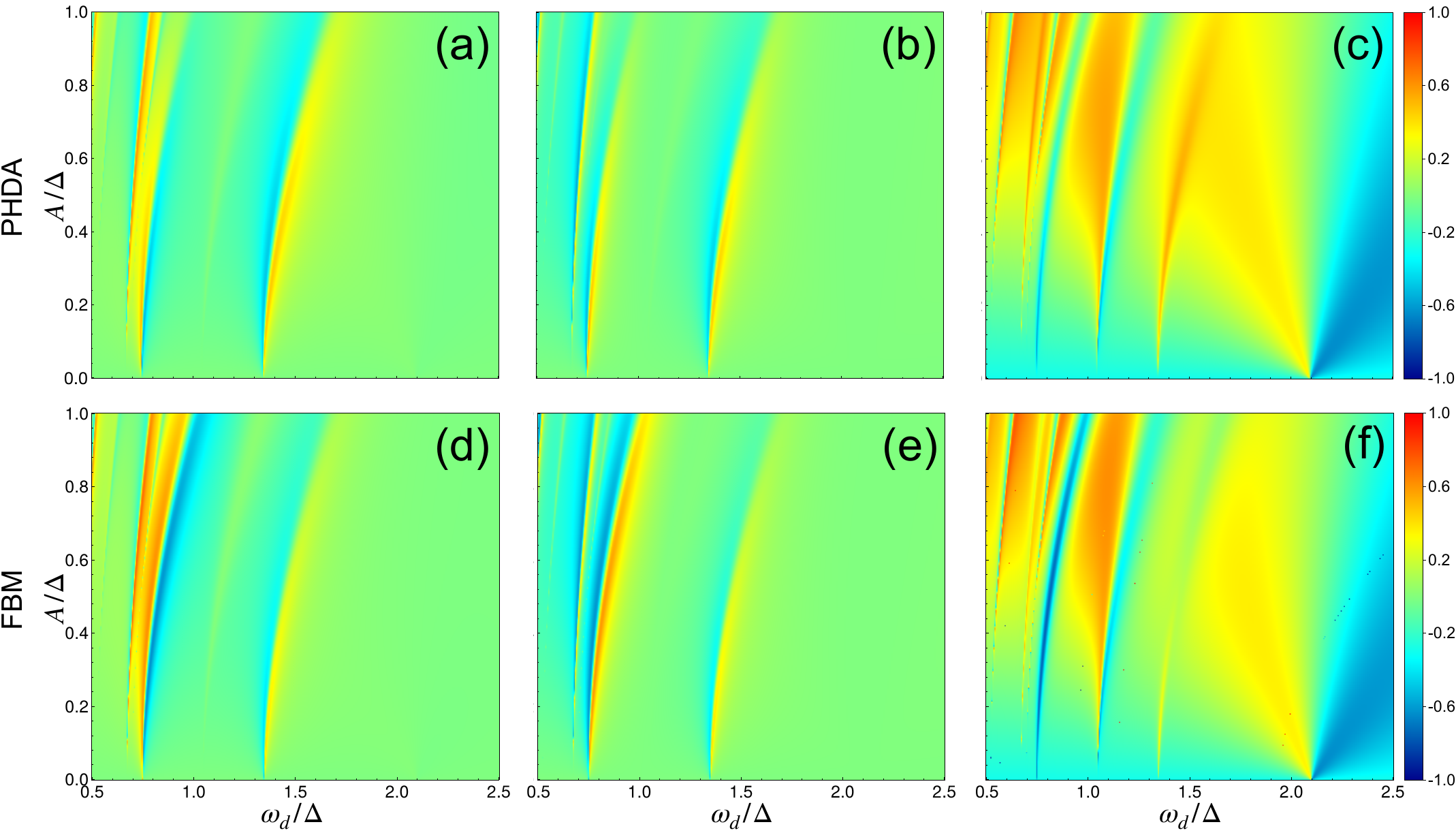}
\caption{\textbf{Phase-resolved coherence-sensitive observables.} All quantities are evaluated at $t_0=\tau/2$. The upper (lower) row corresponds to PHDA (FBM). Panels (a,d), (b,e), and (c,f) display $\mathcal X_1$, $\mathcal X_2$, and the symmetrized mixed quantity $\mathcal X_{12}$, respectively. All panels share the same scale, ranging from $-1$ to $1$, so that differences in the amplitude, width, and shape of the resonant structures can be compared directly. The plotted quantities are system observables, and their relation to calibrated output signals, including the corresponding spectral factors, is given in \Appref{app:floquet}. Parameters are the same as in Fig.~\ref{fig:populations_gap}.}
\label{fig:coherences}
\end{figure*}

By using an overbar on the expectation value to denote the average over one drive period, $\overline{\langle O(t)\rangle}_m\equiv \tau^{-1}\int_0^\tau dt\,\langle O(t)\rangle_m$, where $m$ labels the dissipative construction, and recalling the notation for the field operators of both FBM and PHDA approaches, we calculate the period-averaged dressed populations as

\begin{align}
\overline{\left\langle S_i^+S_i^-\right\rangle}_{\mathrm{PHDA}}
&=\frac{1}{\tau}\int_0^\tau dt\,
\mathrm{Tr}\!\left[S_i^+S_i^-\rho_{\rm pss}^{(\mathrm{PHDA})}(t)\right],\notag\\
\overline{\left\langle \widetilde S_i^+(t)\widetilde S_i^-(t)\right\rangle}_{\mathrm{FBM}}
&=\frac{1}{\tau}\int_0^\tau dt\,
\mathrm{Tr}\!\left[\widetilde S_i^+(t)\widetilde S_i^-(t)\tilde\rho_{\rm ss}\right].
\label{eq:population_general}
\end{align}
For the PHDA the time-independent operator $S_i^+S_i^-$ selects the zeroth Fourier component, and the expression reduces to
\begin{equation}
\overline{\left\langle S_i^+S_i^-\right\rangle}_{\mathrm{PHDA}}
={\rm Tr}\!\left[S_i^+S_i^-\rho_0\right],
\label{eq:population_phda_simplified}
\end{equation}
while for the FBM, the period average and secular channel structure give
\begin{equation}
\overline{\left\langle \widetilde S_i^+(t)\widetilde S_i^-(t)\right\rangle}_{\mathrm{FBM}}
={\rm Tr}\!\left[\sum_{\omega,k}\widetilde S_{i,k}^{+}(\omega)\widetilde S_{i,k}^{-}(\omega) \tilde\rho_{\rm ss}\right].
\label{eq:population_simplified}
\end{equation}
Details of both reductions are reported in \Appref{app:population}. We also diagonalize the vectorized Liouvillian to extract the relaxation gap $\xi=\min_{j\ne0}[-{\rm Re}(\lambda_j)]$, which sets the slowest approach to the asymptotic state, and the explicit superoperator form is given in \Appref{app:gap}.

\section{Population agreement and coherence-level breakdown}
\label{sec:observables_relaxation}

We begin with the benchmark most commonly used in driven spectroscopy, i.e. the period-averaged dressed populations for the PHDA and FBM approaches. Since both methods share the same $H_S(t)$, they inherit the same avoided crossings and multiphoton resonance skeleton. This common coherent structure accounts for the close population agreement over broad regions of weak and moderate modulation, even though the two dissipators select different periodic steady states.

Figure~\ref{fig:populations_gap} shows where that agreement becomes visibly less accurate. The resonance positions remain common to both descriptions, but appreciable differences are mainly evident once the modulation reaches $A/\Delta\gtrsim0.3$. In particular, the FBM population maps develop a stronger asymmetry than PHDA and show a suppression on the left side of the resonant tongue that emerges near $\omega_d/\Delta\simeq2.1$. These features are much less pronounced in the PHDA maps. The relaxation-gap panels display a clearer reorganization of the decay structure. Thus the population comparison supports the conservative conclusion that the two methods reproduce essentially the same resonance skeleton and remain close over a broad parameter domain, but they do not agree quantitatively once the drive itself becomes strong. Conversely, in the low-excitation regime PHDA provides an excellent approximation to the full FBM treatment, as both the dissipative channels and the corresponding system and detection operators closely approximate their Floquet-resolved counterparts, as long as drive-induced dressing remains sufficiently weak.

\begin{figure*}[!t]
\maybeincludegraphics[0.8\textwidth]{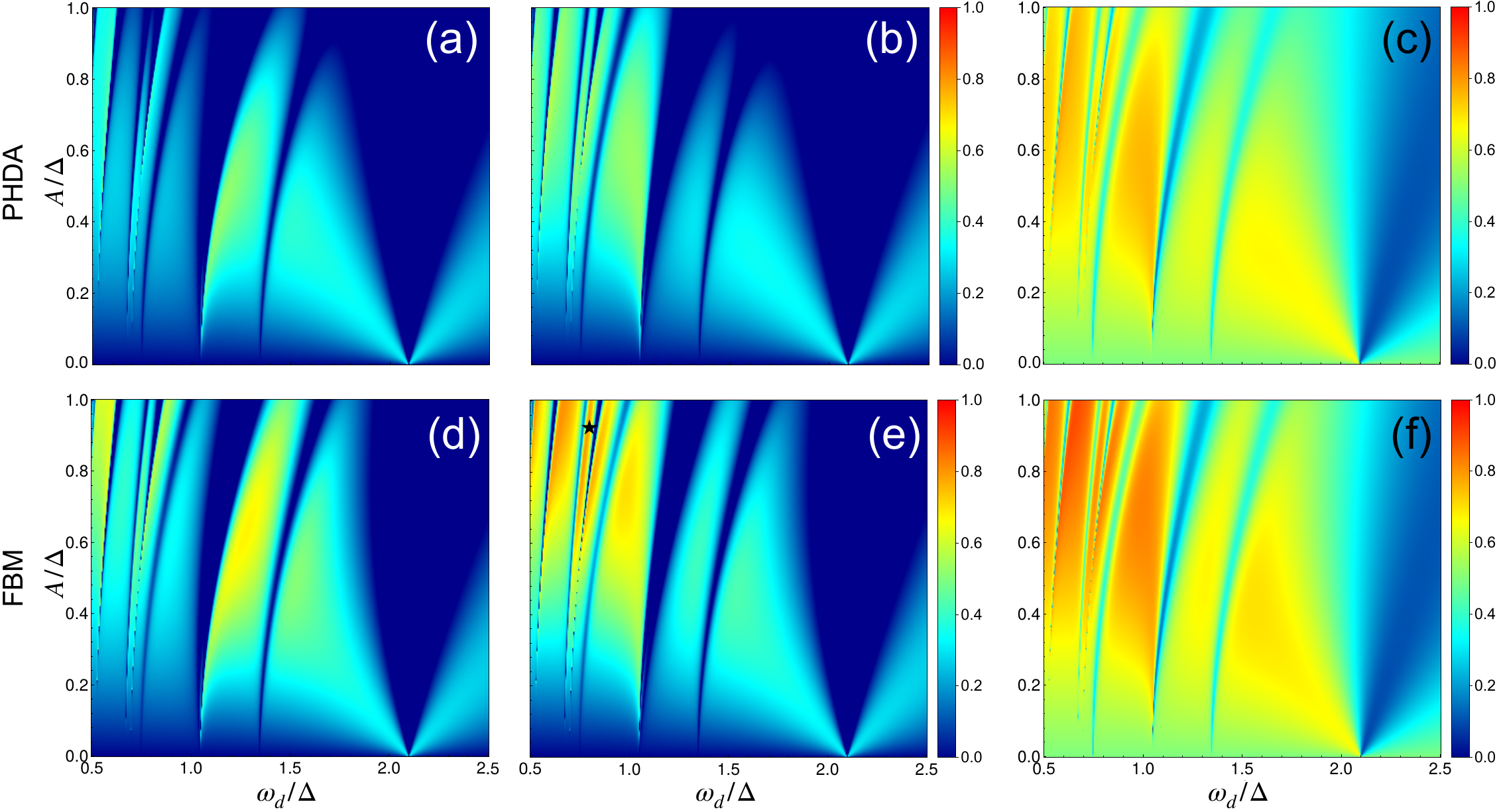}
\caption{\textbf{Concurrence and phase-anchored Floquet fidelity.} The upper row is calculated with PHDA and the lower row with FBM. The left and middle columns show Wootters' concurrence $C(t_0)$ at $t_0/\tau=0$ [panels (a,d)] and $t_0/\tau=1/2$ [panels (b,e)]. The right column [panels (c,f)] shows the overlap $F_{{\rm B},0}(\tau/2)$ with the Floquet Bell-like reference $|\Psi_{\rm B}^{\rm F}(0)\rangle=[|\Phi_1(0)\rangle-|\Phi_4(0)\rangle]/\sqrt2$, constructed at $t=0$ and then held fixed. All scales run from 0 to 1. FBM produces stronger and more extended concurrence domains, especially at half period, and a visibly different fidelity landscape over the same resonant sectors. The black star in panel (e), at $\omega_d/\Delta=0.79$ and $A/\Delta=0.92$, marks the operating point used in Fig.~\ref{fig:finite_temperature}; there $C_{\rm FBM}(\tau/2)\simeq0.8$, close to but below the absolute maximum of panel (e). Parameters are the same as in Fig.~\ref{fig:populations_gap}.}
\label{fig:concurrence_fidelity}
\end{figure*}

To compare the coherence structure predicted by PHDA and FBM approaches, we define
\begin{align*}
\Sigma_i&=S_i^++S_i^-,&
\Sigma_{12}&=\tfrac12\{\Sigma_1,\Sigma_2\},\\
\widetilde\Sigma_i(t)&=\widetilde S_i^+(t)+\widetilde S_i^-(t),&
\widetilde\Sigma_{12}(t)&=\tfrac12\{\widetilde\Sigma_1(t),\widetilde\Sigma_2(t)\}.
\end{align*}
At $t_0=\tau/2$, for $a\in\{1,2,12\}$,
\begin{align}
\mathcal X_a^{(\mathrm{PHDA})}(t_0)&=\mathrm{Tr}\!\left[\Sigma_a\rho_{\rm pss}^{(\mathrm{PHDA})}(t_0)\right],\notag\\
\mathcal X_a^{(\mathrm{FBM})}(t_0)&=\mathrm{Tr}\!\left[\widetilde\Sigma_a(t_0)\tilde\rho_{\rm ss}\right].
\label{eq:coherence_observables}
\end{align}
The anticommutator makes the mixed observable Hermitian and removes an arbitrary operator-ordering choice. In the weak-static-coupling limit and at low excitation density, where the dressed eigenstates remain continuously connected to the uncoupled two-qubit states, these observables include contributions associated with the even-parity coherence between $|gg\rangle$ and $|ee\rangle$ and the exchange coherence between $|ge\rangle$ and $|eg\rangle$. More generally, they quantify coherences in the dressed description and are therefore much more sensitive than a population to the phases and relative weights selected by the dissipative channels.

Figure~\ref{fig:coherences} shows alternating positive and negative lobes separated by zero contours. Their sign reversals track the phase change of the driven response across successive avoided quasienergy crossings, thus a zero contour marks a phase reversal rather than the disappearance of hybridization. The dominant resonances occur in similar locations in both rows, confirming that PHDA still follows the leading coherent Floquet structure. The discrepancy appears mainly as a local enhancement or suppression of the same response and, in several regions, as a broadening or narrowing of the positive and negative lobes. The single-qubit observables show particularly visible changes around the low-frequency resonances, while the mixed observable preserves the large-scale sign pattern but with a redistributed local contrast. Because these are signed observables, changes in channel weights can produce large local differences even where the corresponding populations remain close.
Therefore, Fig.~\ref{fig:coherences} provides a direct coherence-level test of the two dissipative constructions. These intrinsic system coherences are connected to experimentally accessible homodyne signals through the input-output relation. Its PHDA and FBM source terms are evaluated in the same representations used for the respective master equations, with the spectral response of the monitored output line included as described in \Appref{app:floquet}.

\section{Floquet-resolved selection of entanglement}
\label{sec:stroboscopic_concurrence}

\begin{figure*}[!t]
\maybeincludegraphics[1.0\textwidth]{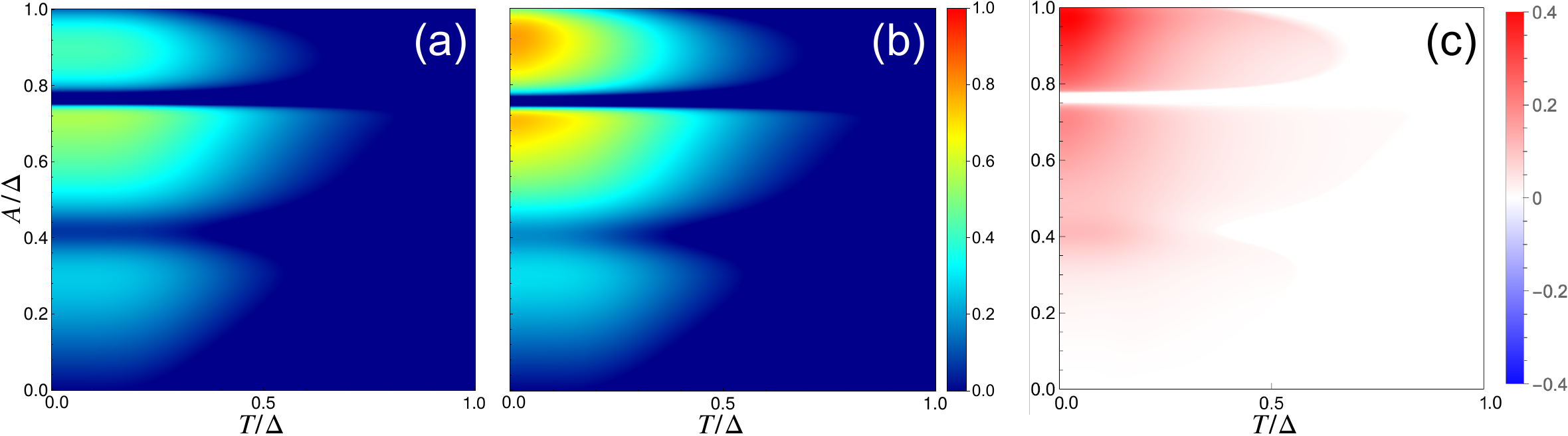}
\caption{\textbf{Thermal robustness at the marked operating frequency.} Half-period concurrence versus reduced bath temperature $T/\Delta$ and modulation amplitude $A/\Delta$, for $T_1=T_2=T$ and $\omega_d/\Delta=0.79$ [star in Fig.~\ref{fig:concurrence_fidelity}(e)]. Panels (a) and (b) show the PHDA and FBM results, respectively, and panel (c) their difference $\Delta C=C_{\rm FBM}-C_{\rm PHDA}$. Color scales are $[0,1]$ in (a,b) and $[-0.4,0.4]$ in (c). At $T/\Delta=0$ and $A/\Delta=0.6$, $C_{\rm FBM}\simeq0.60$ and $\Delta C\simeq0.15$, corresponding to a PHDA underestimate of about $25\%$. Remaining parameters are as in Fig.~\ref{fig:populations_gap}.}
\label{fig:finite_temperature}
\end{figure*}

Concurrence provides a direct comparison between the asymptotic states predicted by the two approaches, since it depends nonlinearly on the full density matrix and is particularly sensitive to differences in its coherence structure. For each dissipative model $m\in{\mathrm{PHDA},\mathrm{FBM}}$, we independently reconstruct the periodic steady state at the same stroboscopic phase $t_0$. In the PHDA, the phase dependence is recovered from the harmonic components as
$\rho_{12}^{(\mathrm{PHDA})}(t_0)=\sum_n\rho_n e^{in\omega_dt_0}$,
whereas in the FBM approach the stationary Floquet interaction-picture solution of Eq.~\eqref{eq:stationary_state} is transformed back according to
$\rho_{12}^{(\mathrm{FBM})}(t_0)=U_S(t_0)\tilde\rho_{\rm ss}U_S^\dagger(t_0)$.
The representation of the density matrix and of the physical two-qubit spin-flip operator is kept consistent throughout.
Entanglement is then quantified using Wootters' concurrence~\cite{Wootters1998}, evaluated directly as $C[\rho_{12}^{(m)}(t_0)]$ for each state. Differences in concurrence therefore reflect differences between the asymptotic states selected by the two dissipative constructions, rather than differences arising solely from the representation of dressed observables. Because concurrence vanishes when the relevant coherences remain below population-dependent separability thresholds, even moderate changes in individual density-matrix elements can move the state across the separability boundary and lead to pronounced differences in the entanglement landscape. The FBM result characterizes the steady state obtained when bath-induced transitions are resolved in the Floquet basis, while the PHDA provides the corresponding prediction based on the static $H_0$ dressed decomposition. Details of the phase-resolved density-matrix reconstruction and concurrence calculation are given in \Appref{app:concurrence}.

The fidelity accompanying the concurrence is evaluated with respect to a phase-anchored Floquet Bell-like reference defined from the common coherent Hamiltonian at $t=0$,
\begin{align}
|\Psi_{\rm B}^{\rm F}(0)\rangle
&\equiv\frac{|\Phi_1(0)\rangle-|\Phi_4(0)\rangle}{\sqrt2},\notag\\
F_{{\rm B},0}^{(m)}(t_0)
&=\langle\Psi_{\rm B}^{\rm F}(0)|\rho_{12}^{(m)}(t_0)|\Psi_{\rm B}^{\rm F}(0)\rangle .
\label{eq:floquet_target_fidelity}
\end{align}
The subscript $0$ indicates that the reference ket is kept fixed at the initial phase, whereas only the density matrix is evaluated at the stroboscopic time $t_0$; the labels 1 and 4 denote the numerical ordering of the Floquet modes. Accordingly, $F_{{\rm B},0}$ is a fixed-reference state-overlap diagnostic, rather than an independent entanglement monotone. In particular, $F_{{\rm B},0}(\tau/2)$ quantifies how closely the half-period state resembles the Bell-like Floquet reference defined at $t=0$.

Figure~\ref{fig:concurrence_fidelity} shows that the apparent agreement of the population maps masks substantial differences in the underlying quantum states. The PHDA approach often keeps the relevant coherences closer to, or below, the separability threshold and therefore predicts weaker and less extended entangled domains, while the FBM approach crosses that threshold over broad portions of the same resonance tongues, with the contrast becoming especially clear at $t_0=\tau/2$. The fidelity panels show that even where both methods identify the same resonant sector, they assign different overlaps with the fixed relative-minus Bell-like reference. The black star in panel (e) selects $\omega_d/\Delta=0.79$ and $A/\Delta=0.92$, where the FBM half-period concurrence is approximately $0.8$. This is a representative high-entanglement point rather than the absolute maximum, and it is used below to test whether the discrepancy persists against both thermal activation and changes in drive amplitude. 

To determine whether this quantum-state discrepancy persists beyond zero temperature, we next examine the neighborhood of the representative operating point marked in Fig.~\ref{fig:concurrence_fidelity}(e). We fix $\omega_d/\Delta=0.79$, set $T_1=T_2=T$, and scan both $T/\Delta$ and $A/\Delta$ while evaluating the concurrence at the half-period phase $t_0=\pi/\omega_d$. With $\hbar=k_{\rm B}=1$, the temperature axis is $T/\Delta$, and restoring physical units gives
\begin{equation}
T_{\rm phys}=\left(\frac{T}{\Delta}\right)\frac{\hbar\Delta}{k_{\rm B}}\simeq 47.99\,\mathrm{mK}\left(\frac{\Delta/2\pi}{\mathrm{GHz}}\right)\left(\frac{T}{\Delta}\right).
\label{eq:temperature_conversion}
\end{equation}
For a representative superconducting-qubit splitting $\Delta/2\pi=5\,\mathrm{GHz}$, the values $T/\Delta=0.1$, $0.5$, and $1$ correspond approximately to $24$, $120$, and $240\,\mathrm{mK}$, respectively. Absorption terms proportional to $n[\Delta_k(\omega),T_i]$ open additional drive-assisted channels and therefore provide a stringent test of the dissipative construction.

Figure~\ref{fig:finite_temperature} shows three entangled amplitude windows that are progressively suppressed as the bath temperature increases. In every window, the FBM map in panel (b) is brighter and extends to higher temperature than the PHDA map in panel (a). The difference map in panel (c) is positive over almost the entire nonzero-concurrence region, showing that PHDA generally underestimates the concurrence rather than merely displacing an isolated resonance. A representative quantitative example occurs at $T/\Delta=0$ and $A/\Delta=0.6$, where $C_{\rm FBM}\simeq0.60$ and $C_{\rm FBM}-C_{\rm PHDA}=\Delta C\simeq0.15$. Thus the PHDA approach underestimates the FBM prediction by approximately $\Delta C/C_{\rm FBM}\simeq25\%$. Thermal activation therefore does not simply wash out a common state at two slightly different rates. The additional absorption channels are weighted differently by the static and Floquet-resolved dissipators, and the resulting concurrence deficit remains substantial throughout the robust entangled bands.

\section{Discussion and outlook}

A central message of this work is that agreement at the level of populations and resonance positions does not, by itself, guarantee agreement on the underlying quantum state. In our parametrically coupled qubits, the common coherent Floquet structure is sufficient to make the PHDA and FBM population maps nearly indistinguishable over broad regions of parameter space. The distinction becomes progressively more pronounced, however, for observables that retain phase information: differences first emerge at the level of dressed coherences and ultimately translate into extended regions of entanglement predicted by the FBM treatment that are generally underestimated by PHDA.

The comparison also clarifies the origin of the discrepancies between the two approaches. The coherent Hamiltonian, reservoirs, microscopic system--bath couplings, and physical observables are the same; the distinction lies in how the driven dynamics is incorporated into the dissipative description. PHDA retains the full periodic coherent evolution but resolves bath-induced transitions in the static dressed basis of $H_0$, whereas FBM decomposes the same coupling operators between Floquet modes and their associated sidebands, consistently carrying this structure into the operators entering the input--output relations. Within the common weak-coupling system--bath framework adopted here, the numerically converged FBM treatment therefore serves as the reference against which the PHDA approximation is assessed.
Periodic driving redistributes static transitions among Floquet-assisted dissipative channels that are not individually resolved in PHDA. As a result, the main resonance structure remains largely unchanged, while increasingly phase-sensitive quantities exhibit progressively larger deviations: the relaxation gap shows an intermediate response, whereas single-qubit and mixed coherences are affected more strongly. Concurrence, evaluated for the two periodic steady states at the same stroboscopic phase, translates these coherence differences into a pronounced change in entanglement, while the fixed-reference fidelity provides a complementary measure of the resulting state composition. The finite-temperature analysis further shows that this discrepancy persists when thermally activated channels become accessible.

Our conclusions do not rely on applying full secularization outside its range of validity. The FBM generator is of GKLS form and yields completely positive, trace-preserving dynamics in the weak-damping regime considered here. The explicit channel-resolution map shows that unresolved transitions are restricted to narrow near-degeneracy loci and do not appreciably overlap the broad coherence and entanglement domains. Nonsecular or partial-secular Floquet theories remain essential inside those loci and in more strongly broadened systems~\cite{Akbari2026,Farina2019,NathanRudner2020,Mozgunov2020,Trushechkin2021}; in the present parameter window they would refine local features rather than the main state-selection result.

The practical implication is broader than the two-qubit model. Population spectroscopy is a necessary but insufficient benchmark whenever the target resource depends on coherences, including entanglement, fidelity, and phase-sensitive correlations. A robust validation strategy should therefore combine diagonal observables with at least one coherence-sensitive or state-level diagnostic. In larger driven arrays, where many sideband-resolved pathways compete, this requirement should become even more consequential. Parametrically coupled superconducting circuits provide a natural experimental platform for testing this hierarchy and, at the same time, point toward the broader possibility of harnessing Floquet-resolved dissipation as a resource for tailoring nonequilibrium quantum states.

\paragraph*{Author contributions.}
A.R. conceived and developed the theory and supervised the project. R.N. and G.P. developed and ran the numerical code and produced the figures. All authors analyzed and interpreted the results and contributed to writing and revising the manuscript.

\begin{acknowledgments}
We thank Dr. Giuseppe Cassone for fruitful discussions. G.P. acknowledges financial support from the Centro Siciliano di Fisica Nucleare e Struttura della Materia (CSFNSM). The authors thank OpenAI ChatGPT (GPT-5.6 Sol) for assistance with the organization, revision, and refinement of portions of the manuscript. All scientific arguments, calculations, results, and conclusions were established, checked, and validated by the authors, who take full responsibility for the content.
\end{acknowledgments}

\section*{Data availability}
The numerical data and source code that support the findings of this article are available from the corresponding author upon reasonable request.

\appendix

\section{Static-dressed harmonic and Floquet-resolved constructions}
\label{app:floquet}

This appendix collects the implementation details not given explicitly in Sec.~\ref{sec:model}. The common microscopic Hamiltonian, the static-dressed operators $S_{i,\nu}^{\pm}$ used in PHDA, and the Floquet-resolved operators $\widetilde S_{i,k}^{\pm}(\omega)$ used in FBM are defined in Eqs.~\eqref{eq:total_hamiltonian}, \eqref{eq:static_dressed_operators}, and \eqref{eq:floquet_operator_decomp}--\eqref{eq:fbm_master}. The two approaches therefore share the same physical couplings and observables; what changes is their dynamical representation. Here we give the PHDA harmonic-space construction and the interaction-picture form from which the FBM dissipator follows.

For PHDA, all matrices are expressed in the fixed static-dressed representation introduced in the main text. The driven Hamiltonian reads
\begin{equation}
\begin{aligned}
H_{S,\rm d}(t)&=H_{0,\rm d}+\sum_{q=\pm1}H_{q,\rm d}e^{iq\omega_dt},\\
H_{+1,\rm d}&=H_{-1,\rm d}=\frac{A}{2}U_0^\dagger\sigma_x^{(1)}\sigma_x^{(2)}U_0 ,
\end{aligned}
\end{equation}
and the PHDA master equation is
\begin{align}
\dot\rho(t)&=\mathcal L_{0,\rm d}^{\rm PHDA}\rho(t)
-i\sum_{q=\pm1}e^{iq\omega_dt}[H_{q,\rm d},\rho(t)],\notag\\
\mathcal L_{0,\rm d}^{\rm PHDA}\rho&=-i[H_{0,\rm d},\rho]
+\mathcal D_{\rm PHDA}^{\rm d}\rho ,
\label{eq:sm_phda_time}
\end{align}
with
\begin{align}
\mathcal D_{\rm PHDA}^{\rm d}\rho
={}&\sum_{i,\nu>0}\frac{\gamma^{(i)}(\nu)}{2}
\{1+n(\nu,T_i)\}\mathcal D[S_{i,\nu}^{-}]\rho\notag\\
&+\sum_{i,\nu>0}\frac{\gamma^{(i)}(\nu)}{2}
n(\nu,T_i)\mathcal D[S_{i,\nu}^{+}]\rho .
\label{eq:sm_phda_dissipator}
\end{align}
The full coherent modulation is retained by expanding the periodic steady state as
\begin{equation}
\rho_{\rm pss}^{(\rm PHDA)}(t)=\sum_{n=-\infty}^{\infty}\rho_n e^{in\omega_dt},
\end{equation}
which gives
\begin{equation}
0=\left(\mathcal L_{0,\rm d}^{\rm PHDA}-in\omega_d\right)\rho_n
-i\sum_{q=\pm1}[H_{q,\rm d},\rho_{n-q}].
\label{eq:sm_phda_harmonics}
\end{equation}
We impose ${\rm Tr}\rho_0=1$ and ${\rm Tr}\rho_{n\neq0}=0$ and truncate the Fourier ladder symmetrically until the observables converge~\cite{Mercurio2026DPT}. Thus the drive enters the PHDA dynamics nonperturbatively through the coupled harmonics, while its dissipative channels remain those of the static dressed Hamiltonian $H_0$.

For FBM, the $H_0$ eigenbasis is only a numerical representation: the coupling operators are resolved after solving the complete Floquet problem. Using the interaction picture generated by $U_S(t)\otimes e^{-iH_Bt}$ and the decomposition of Eq.~\eqref{eq:floquet_operator_decomp}, the common weak-coupling RWA retains only excitation-exchange terms. With $\delta_{nk}^{(i)}(\omega)=\omega_n^{(i)}-\Delta_k(\omega)$,
\begin{align}
\tilde H_{SB}^{\rm RWA}(t)=&\sum_{n,\omega,k}g_n^{(1)}
 b_ne^{-i\delta_{nk}^{(1)}(\omega)t}\widetilde S_{1,k}^{+}(\omega)\notag\\
&+\sum_{n,\omega,k}g_n^{(1)}
 b_n^\dagger e^{i\delta_{nk}^{(1)}(\omega)t}\widetilde S_{1,k}^{-}(\omega)\notag\\
&+\sum_{n,\omega,k}g_n^{(2)}
 c_ne^{-i\delta_{nk}^{(2)}(\omega)t}\widetilde S_{2,k}^{+}(\omega)\notag\\
&+\sum_{n,\omega,k}g_n^{(2)}
 c_n^\dagger e^{i\delta_{nk}^{(2)}(\omega)t}\widetilde S_{2,k}^{-}(\omega).
\label{eq:sm_floquet_interaction}
\end{align}
The $(\omega,k)$ labels retain the total transition frequencies $\Delta_k(\omega)$ required by the subsequent Born--Markov and channel-secular construction. Applying the secularization specified in Sec.~\ref{sec:model}, and following closely the calculations developed in Ref.~\cite{Bluemel1991}, yields Eq.~\eqref{eq:fbm_master}, with transitions sharing the same $\omega$ at fixed $k$ already grouped in $X_k^{(i)}(\omega)$. The Lamb-shift contribution is neglected in both treatments.

\paragraph{Dressed input-output relations.}

Consistently with the operator-action convention introduced in the main text, $B_{i,\rm in/out}^{-}$ and $B_{i,\rm in/out}^{+}=[B_{i,\rm in/out}^{-}]^\dagger$ denote, respectively, the annihilation and creation components of the input/output bath fields. This notation should not be confused with the alternative optical convention in which the positive-frequency field is commonly denoted by a superscript $(+)$. Assuming weak coupling between each qubit and its corresponding input--output bath channel, within the Markov approximation, the usual input--output relations can be obtained in the Heisenberg picture,
\begin{equation}
B_{i,\rm out}^{-}(t)=B_{i,\rm in}^{-}(t)+\mathcal J_i^{-}(t),
\label{eq:sm_input_output}
\end{equation}
while the corresponding relation for $B_{i,\rm out}^{+}(t)$ follows by Hermitian conjugation. The system source term $\mathcal J_i^{-}(t)$ is built from the dressed lowering transitions coupled to the monitored line~\cite{GardinerCollett1985,CiutiCarusotto2006,Ridolfo2012,Garziano2017,Settineri2021,Macri2022}.

To keep the dissipative reservoir and the readout response conceptually distinct, let $\Gamma_i^{\rm det}(\nu)$ denote the coupling spectrum of the detection line. If this is the same physical channel that produces the loss appearing in the master equation, a convenient parametrization is $\Gamma_i^{\rm det}(\nu)=\eta_i(\nu)\gamma^{(i)}(\nu)$, where $\eta_i(\nu)$ is a dimensionless collection/detector transfer function. For an ideal broadband detector $\eta_i$ is approximately constant, and in the common ideal case $\eta_i=1$ the standard coefficient $\sqrt{\gamma^{(i)}(\nu)}$ is recovered. If the readout line is independent of the dissipative reservoir, $\Gamma_i^{\rm det}(\nu)$ is instead an independent spectral function.

For the evaluation of output observables it is convenient to represent the source operator in the same picture in which the corresponding master equation is solved. In PHDA we therefore use its Schr\"odinger-picture form in the fixed static-dressed representation, whereas in FBM we use the Floquet interaction-picture form. Resolving the detected field channel by channel gives
\begin{align}
\mathcal J_{i,\rm PHDA}^{-}
&=\sum_{\nu>0}\sqrt{\Gamma_i^{\rm det}(\nu)}\,S_{i,\nu}^{-},\notag\\
\widetilde{\mathcal J}_{i,\rm FBM}^{-}(t)
&=\sum_{\omega,k}\sqrt{\Gamma_i^{\rm det}[\Delta_k(\omega)]}\,
e^{-i\Delta_k(\omega)t}\widetilde S_{i,k}^{-}(\omega),
\label{eq:sm_detection_operators}
\end{align}
and these are the PHDA and FBM representations of the system source entering Eq.~\eqref{eq:sm_input_output}. Accordingly, expectation values and correlation functions of the source contribution can be evaluated directly with $\rho_{\rm pss}^{(\rm PHDA)}(t)$ in PHDA and with $\tilde\rho_{\rm ss}$ in FBM, as the expectation values are frame independent. 
The main-text population and coherence maps are intentionally reported as system observables, without the detector-dependent factors $\Gamma_i^{\rm det}$. To compare them with calibrated emission rates or homodyne currents, the spectral weights in Eq.~\eqref{eq:sm_detection_operators} must be included. The homodyne quadrature is
\begin{equation}
I_{i,\theta}(t)\propto e^{-i\theta}B_{i,\rm out}^{-}(t)
+e^{i\theta}B_{i,\rm out}^{+}(t),
\end{equation}
where $\theta$ represents the phase of the local oscillator. For vacuum input, normally ordered output moments reduce to moments of the corresponding dressed system operators multiplied by the appropriate channel weights. Thus, the reservoir spectrum and the detector response coincide only when the same physical line provides both dissipation and readout; otherwise, they represent distinct experimental inputs~\cite{GardinerCollett1985,Clerk2010}.

\begin{figure*}[!t]
\maybeincludegraphics[0.44\textwidth]{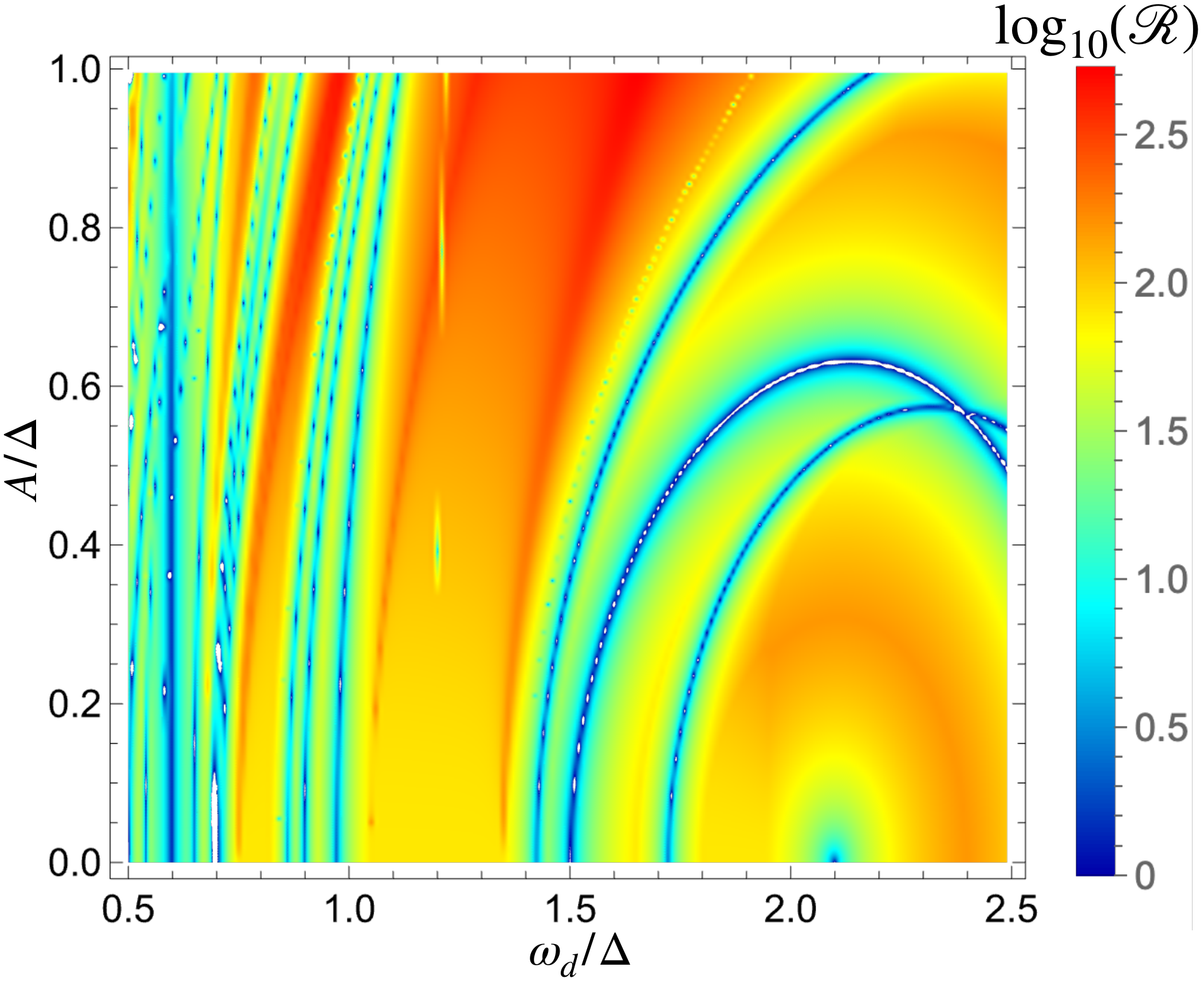}
\caption{\textbf{Quantitative validity of the secular approximation.} Minimum channel-resolution ratio $\mathcal R_{\min}$, plotted as $\log_{10}(\mathcal R_{\min})$ in the $(\omega_d/\Delta,A/\Delta)$ plane. The parameters are $\Delta_1=\Delta_2=\Delta$, $\epsilon_1/\Delta=0.1$, $\epsilon_2=0$, $g_0/\Delta=0.3$, $T_1=T_2=0$, and $\gamma/\Delta=10^{-3}$; the linewidths are evaluated with the same Ohmic reservoir spectral density used in the FBM dissipator. The color scale starts at $\log_{10}(\mathcal R_{\min})=0$, corresponding to the linear threshold $\mathcal R_{\min}=1$. Values below unity are clipped and shown in white. The white set is confined to extremely narrow near-degeneracy loci, while essentially the entire parameter plane satisfies the secular-resolution condition and most regions lie well above threshold.}
\label{fig:secularity}
\end{figure*}

\section{Floquet transition rates and Pauli population equation}
\label{app:rates}
The channel-resolved Floquet transition rates entering the Pauli population equation, Eq.~\eqref{eq:pauli_population}, are specified below. The corresponding rate matrix satisfies
\begin{equation}
M_{\alpha\beta}=
\begin{cases}
W_{\alpha\leftarrow\beta}, & \alpha\ne\beta,\\[4pt]
-\sum_\gamma W_{\gamma\leftarrow\alpha}, & \alpha=\beta,
\end{cases}
\end{equation}

and the stationary populations are obtained from $M\mathbf p^{\rm ss}=0$ with $\sum_\alpha p_\alpha^{\rm ss}=1$.
For the present problem the number of emitters is $N_{\rm at}=2$. The total transition rate from the Floquet state $\beta$ to the Floquet state $\alpha$ is a sum over emitters, quasienergy differences, and drive harmonics,
\begin{equation}
W_{\alpha\leftarrow\beta}=
\sum_{i=1}^{N_{\rm at}}\sum_\omega\sum_k
\left[W_{\alpha\leftarrow\beta}^{(i,+)}(\omega,k)+W_{\alpha\leftarrow\beta}^{(i,-)}(\omega,k)\right].
\label{eq:sm_total_rate}
\end{equation}
We define the channel matrix elements
\begin{equation}
\mathcal M_{\alpha\beta}^{(i,\pm)}(\omega,k)
=\langle\Phi_\alpha(0)|\widetilde S_{i,k}^{\pm}(\omega)|\Phi_\beta(0)\rangle .
\end{equation}
The absorption-like and emission-like contributions are then
\begin{align}
W_{\alpha\leftarrow\beta}^{(i,+)}(\omega,k)&=
\gamma^{(i)}[\Delta_k(\omega)]\,n[\Delta_k(\omega),T_i]\notag\\
&\quad\times\left|\mathcal M_{\alpha\beta}^{(i,+)}(\omega,k)\right|^2,
\label{eq:sm_rate_plus}\\
W_{\alpha\leftarrow\beta}^{(i,-)}(\omega,k)&=
\gamma^{(i)}[\Delta_k(\omega)]\left\{1+n[\Delta_k(\omega),T_i]\right\}\notag\\
&\quad\times\left|\mathcal M_{\alpha\beta}^{(i,-)}(\omega,k)\right|^2.
\label{eq:sm_rate_minus}
\end{align}
Using the explicit Floquet decomposition of the dressed operators, Eq.~\eqref{eq:sm_total_rate} can be written as
\begin{align}
W_{\alpha\leftarrow\beta}
=&\sum_{i=1}^{N_{\rm at}}\sum_\omega\sum_k
\gamma^{(i)}[\Delta_k(\omega)]\Bigl[
 n[\Delta_k(\omega),T_i]\Theta[\Delta_k(\omega)]\notag\\
&\quad\times |X_{\alpha\beta}^{(i)}(k)|^2
\delta_{\omega,\mu_\alpha-\mu_\beta}\notag\\
&+\{1+n[\Delta_k(\omega),T_i]\}\Theta[\Delta_k(\omega)]\notag\\
&\quad\times |X_{\beta\alpha}^{(i)}(k)|^2
\delta_{\omega,\mu_\beta-\mu_\alpha}\Bigr].
\label{eq:sm_rate_x}
\end{align}
Equivalently, one may define the partial rate at a fixed quasienergy difference,
\begin{align}
W_{\alpha\leftarrow\beta}(\omega)
=&\sum_{i=1}^{N_{\rm at}}\sum_k
\gamma^{(i)}[\Delta_k(\omega)]\Bigl[
 n[\Delta_k(\omega),T_i]\notag\\
&\quad\times\left|\mathcal M_{\alpha\beta}^{(i,+)}(\omega,k)\right|^2\notag\\
&+\{1+n[\Delta_k(\omega),T_i]\}\notag\\
&\quad\times\left|\mathcal M_{\alpha\beta}^{(i,-)}(\omega,k)\right|^2
\Bigr].
\end{align}
so that $W_{\alpha\leftarrow\beta}=\sum_\omega W_{\alpha\leftarrow\beta}(\omega)$. Projecting the fully secular FBM equation onto the diagonal Floquet populations,
$p_\alpha=\langle\Phi_\alpha(0)|\tilde\rho|\Phi_\alpha(0)\rangle$,
gives the Pauli equation whenever the diagonal and off-diagonal sectors are dynamically decoupled, as in the standard nondegenerate fully secular Floquet limit~\cite{Kohler1997,Grifoni1998,Hone2009}.

If exact or unresolved degeneracies prevent the decoupling of populations and coherences, the Pauli rate equation is no longer closed and the stationary state must be obtained from a master equation retaining the corresponding coherences, such as the full Liouvillian equation used in Refs.~\cite{Kohler1997,Hone2009}.

\section{Quantitative secular-resolution test}
\label{app:secularity}
To quantify the spectral resolution underlying the fully secular generator, we label each retained Floquet transition channel by $a$ and denote its total physical transition frequency by $\nu_a=\Delta_{k_a}(\omega_a)$. For each pair of distinct channels, the separation is compared with the sum of their dissipative linewidths,
\begin{equation}
\mathcal R_{ab}=\frac{|\nu_a-\nu_b|}{\Gamma_a+\Gamma_b},\qquad
\mathcal R_{\min}=\min_{a\neq b}\mathcal R_{ab}.
\label{eq:sm_secular_ratio}
\end{equation}
The minimum is taken over the complete ratio for all distinct retained channel pairs. For a channel $a=(\omega_a,k_a)$ with total transition frequency $\nu_a=\Delta_{k_a}(\omega_a)$, the corresponding dissipative linewidth at zero temperature is evaluated as $\Gamma_a= \sum_i \gamma^{(i)}(\nu_a) \sum_{\mu_\alpha-\mu_\beta=\omega_a} \left|X_{\alpha\beta}^{(i)}(k_a)\right|^2$,
with an analogous definition for $\Gamma_b$. Thus, each linewidth includes the Ohmic reservoir spectral weight together with the strengths of all transitions grouped within the corresponding $(\omega,k)$ channel. 
The threshold $\mathcal R_{\min}=1$ marks equality between the smallest channel separation and the associated linewidth sum, while $\mathcal R_{\min}\gg1$ identifies a well-resolved secular regime.
Figure~\ref{fig:secularity} reports $\log_{10}(\mathcal R_{\min})$ over the full $(\omega_d/\Delta,A/\Delta)$ window explored in the main text. The parameters are $\Delta_1=\Delta_2=\Delta$, $\epsilon_1/\Delta=0.1$, $\epsilon_2=0$, $g_0/\Delta=0.3$, $T_1=T_2=0$, and $\gamma/\Delta=10^{-3}$, with the same Ohmic spectral density used in the FBM dissipator. The color scale begins at zero, corresponding to $\mathcal R_{\min}=1$. Points with $\mathcal R_{\min}<1$ were clipped and rendered white. These unresolved points occupy only extremely narrow near-degeneracy loci, whereas the overwhelming majority of the plane has $\mathcal R_{\min}>1$ and broad regions satisfy $\mathcal R_{\min}\gg1$. This direct test establishes that the population, coherence, concurrence, fidelity, and thermal-robustness maps are evaluated in a quantitatively well-resolved secular regime.

\section{Period-averaged dressed populations}
\label{app:population}
For PHDA, Eq.~\eqref{eq:population_general} contains a time-independent $H_0$-dressed operator, so the period-averaged expectation value selects only the zeroth harmonic:
\begin{equation}
\overline{\left\langle S_i^+S_i^-\right\rangle}_{\mathrm{PHDA}}
={\rm Tr}\!\left[S_i^+S_i^-\rho_0\right],
\end{equation}
which is Eq.~\eqref{eq:population_phda_simplified}.
For FBM, we prove Eq.~\eqref{eq:population_simplified}. For a pair of Hermitian-conjugate Floquet-dressed operators,
\begin{equation}
\widetilde S_i^{+}(t)\widetilde S_i^{-}(t)=\sum_{\omega,k}\sum_{\omega',k'}e^{i[\Delta_k(\omega)-\Delta_{k'}(\omega')]t}\widetilde S_{i,k}^{+}(\omega)\widetilde S_{i,k'}^{-}(\omega').
\end{equation}
Taking the stationary expectation value $\langle\cdots\rangle_{\rm ss}\equiv{\rm Tr}(\cdots\,\tilde\rho_{\rm ss})$ gives the same double sum. Under the secular channel structure, the trace vanishes unless the two operators connect Floquet states separated by the same quasienergy difference, $\omega'=\omega$. The remaining phase is $e^{i(k-k')\omega_dt}$, and the period average gives
\begin{equation}
\frac{1}{\tau}\int_0^\tau dt\,e^{i(k-k')\omega_dt}=\delta_{kk'}.
\end{equation}
All off-diagonal harmonic contributions therefore vanish, yielding Eq.~\eqref{eq:population_simplified}.

\section{Liouvillian gap}
\label{app:gap}

The Liouvillian gap quantifies the slowest exponential approach to the asymptotic periodic state and is therefore the relaxation diagnostic reported in Fig.~\ref{fig:populations_gap}~\cite{Macieszczak2016,Minganti2018}. We vectorize the relevant time-independent generator using ${\rm vec}(A\rho B)=(B^T\otimes A){\rm vec}(\rho)$. For FBM, $\mathbb L$ is the vectorized generator of Eq.~\eqref{eq:fbm_master}, and its spectrum can be used directly. For PHDA, $\mathbb L$ is instead the truncated static-dressed harmonic-space block operator defined by Eq.~\eqref{eq:sm_phda_harmonics}. In this extended Floquet-Liouville representation, eigenvalues differing by integer multiples of $i\omega_d$ correspond to equivalent harmonic replicas; we therefore select one representative within the first Brillouin zone, $-\omega_d/2<{\rm Im}(\lambda_j)\leq\omega_d/2$. The Liouvillian gap is then defined in both cases as
\begin{equation}
\xi=\min_{j\ne0}\left[-{\rm Re}(\lambda_j)\right],
\end{equation}
where the stationary mode $\lambda_0=0$ is excluded. In the figures we report the scaled gap $\xi/\gamma$.

\section{Concurrence and fixed-reference Floquet fidelity}
\label{app:concurrence}
For PHDA, the trace-normalized PSS at phase $t_0$ is obtained by restoring the Fourier phases of the harmonic components,
\begin{equation}
\rho_{12}^{(\mathrm{PHDA})}(t_0)=
\sum_n\rho_n e^{in\omega_dt_0}.
\label{eq:phda_state_reconstruction}
\end{equation}
For FBM, the trace-normalized stationary interaction-picture solution is reconstructed at the same phase through the driven propagator,
\begin{equation}
\rho_{12}^{(\mathrm{FBM})}(t_0)=U_S(t_0)\tilde\rho_{\rm ss}^{(\mathrm{FBM})}U_S^\dagger(t_0).
\label{eq:fbm_state_reconstruction}
\end{equation}
The Fourier factors in Eq.~\eqref{eq:phda_state_reconstruction} and the propagator in Eq.~\eqref{eq:fbm_state_reconstruction} restore the physical stroboscopic phase in the two respective constructions.
The concurrence is subsequently evaluated in the same fixed two-qubit representation, using Wootters' spin-flip construction~\cite{Wootters1998}. For both dissipative descriptions, the phase-resolved density matrix is expressed in the same fixed two-qubit representation, which is kept unchanged throughout the numerical evaluation. In this representation we define $\mathcal Y=\sigma_y\otimes\sigma_y$, with complex conjugation taken in the same representation. The spin-flipped product is then
\begin{equation}
R=\rho_{12}^{(m)}(t_0)\,\mathcal Y\,
[\rho_{12}^{(m)}(t_0)]^*\,\mathcal Y ,
\end{equation}
and, denoting by $\lambda_1\ge\lambda_2\ge\lambda_3\ge\lambda_4$ the eigenvalues of $R$, the concurrence is
\begin{equation}
C=\max\left[0,\sqrt{\lambda_1}-\sqrt{\lambda_2}-\sqrt{\lambda_3}-\sqrt{\lambda_4}\right].
\end{equation}

The fidelity in Fig.~\ref{fig:concurrence_fidelity} uses the Floquet Bell-like reference $|\Psi_{\rm B}^{\rm F}(0)\rangle=\dfrac{1}{\sqrt2}(|\Phi_1(0)\rangle-|\Phi_4(0)\rangle)$ fixed at the initial phase,
and is evaluated as
\begin{equation}
F_{{\rm B},0}^{(m)}(t_0)=\langle\Psi_{\rm B}^{\rm F}(0)|\rho_{12}^{(m)}(t_0)|\Psi_{\rm B}^{\rm F}(0)\rangle.
\end{equation}
The same Floquet-mode ordering and phase convention is used to construct the reference at $t=0$ throughout the numerical evaluation.

\bibliography{bibliografia2}

@article{Reagor2018,
  author  = {M. Reagor and others},
  title   = {Demonstration of universal parametric entangling gates on a multi-qubit lattice},
  journal = {Sci. Adv.},
  volume  = {4},
  pages   = {eaao3603},
  year    = {2018},
  doi     = {10.1126/sciadv.aao3603}
}

@article{Hong2020,
  author  = {S. S. Hong and others},
  title   = {Demonstration of a parametrically activated entangling gate protected from flux noise},
  journal = {Phys. Rev. A},
  volume  = {101},
  pages   = {012302},
  year    = {2020},
  doi     = {10.1103/PhysRevA.101.012302}
}

@article{Sung2021,
  author  = {Y. Sung and others},
  title   = {Realization of high-fidelity {CZ} and {$ZZ$}-free {iSWAP} gates with a tunable coupler},
  journal = {Phys. Rev. X},
  volume  = {11},
  pages   = {021058},
  year    = {2021},
  doi     = {10.1103/PhysRevX.11.021058}
}

@article{McKay2016,
  author  = {D. C. McKay and S. Filipp and A. Mezzacapo and E. Magesan and J. M. Chow and J. M. Gambetta},
  title   = {Universal gate for fixed-frequency qubits via a tunable bus},
  journal = {Phys. Rev. Applied},
  volume  = {6},
  pages   = {064007},
  year    = {2016},
  doi     = {10.1103/PhysRevApplied.6.064007}
}

@article{Roth2017,
  author  = {M. Roth and M. Ganzhorn and N. Moll and S. Filipp and G. Salis and S. Schmidt},
  title   = {Analysis of a parametrically driven exchange-type gate and a two-photon excitation gate between superconducting qubits},
  journal = {Phys. Rev. A},
  volume  = {96},
  pages   = {062323},
  year    = {2017},
  doi     = {10.1103/PhysRevA.96.062323},
  note    = {Erratum: Phys. Rev. A 97, 049903 (2018)}
}

@article{Caldwell2018,
  author  = {S. A. Caldwell and others},
  title   = {Parametrically activated entangling gates using transmon qubits},
  journal = {Phys. Rev. Applied},
  volume  = {10},
  pages   = {034050},
  year    = {2018},
  doi     = {10.1103/PhysRevApplied.10.034050}
}

@article{Ganzhorn2020,
  author  = {M. Ganzhorn and G. Salis and D. J. Egger and A. Fuhrer and M. Mergenthaler and C. M{\"u}ller and P. M{\"u}ller and S. Paredes and M. Pechal and M. Werninghaus and S. Filipp},
  title   = {Benchmarking the noise sensitivity of different parametric two-qubit gates in a single superconducting quantum computing platform},
  journal = {Phys. Rev. Research},
  volume  = {2},
  pages   = {033447},
  year    = {2020},
  doi     = {10.1103/PhysRevResearch.2.033447}
}

@article{Sete2021,
  author  = {E. A. Sete and N. Didier and A. Q. Chen and S. Kulshreshtha and R. Manenti and S. Poletto},
  title   = {Parametric-resonance entangling gates with a tunable coupler},
  journal = {Phys. Rev. Applied},
  volume  = {16},
  pages   = {024050},
  year    = {2021},
  doi     = {10.1103/PhysRevApplied.16.024050}
}

@article{Shirley1965,
  author  = {J. H. Shirley},
  title   = {Solution of the {Schr\"odinger} equation with a Hamiltonian periodic in time},
  journal = {Phys. Rev.},
  volume  = {138},
  pages   = {B979--B987},
  year    = {1965},
  doi     = {10.1103/PhysRev.138.B979}
}

@article{Sambe1973,
  author  = {H. Sambe},
  title   = {Steady states and quasienergies of a quantum-mechanical system in an oscillating field},
  journal = {Phys. Rev. A},
  volume  = {7},
  pages   = {2203--2213},
  year    = {1973},
  doi     = {10.1103/PhysRevA.7.2203}
}

@article{Grifoni1998,
  author  = {M. Grifoni and P. H{\"a}nggi},
  title   = {Driven quantum tunneling},
  journal = {Phys. Rep.},
  volume  = {304},
  pages   = {229--354},
  year    = {1998},
  doi     = {10.1016/S0370-1573(98)00022-2}
}

@article{Bukov2015,
  author  = {M. Bukov and L. D'Alessio and A. Polkovnikov},
  title   = {Universal high-frequency behavior of periodically driven systems: From dynamical stabilization to {Floquet} engineering},
  journal = {Adv. Phys.},
  volume  = {64},
  pages   = {139--226},
  year    = {2015},
  doi     = {10.1080/00018732.2015.1055918}
}

@article{Silveri2017,
  author  = {M. P. Silveri and J. A. Tuorila and E. V. Thuneberg and G. S. Paraoanu},
  title   = {Quantum systems under frequency modulation},
  journal = {Rep. Prog. Phys.},
  volume  = {80},
  pages   = {056002},
  year    = {2017},
  doi     = {10.1088/1361-6633/aa5170}
}

@article{Eckardt2017,
  author  = {A. Eckardt},
  title   = {Colloquium: Atomic quantum gases in periodically driven optical lattices},
  journal = {Rev. Mod. Phys.},
  volume  = {89},
  pages   = {011004},
  year    = {2017},
  doi     = {10.1103/RevModPhys.89.011004}
}

@article{Oka2019,
  author  = {T. Oka and S. Kitamura},
  title   = {{Floquet} engineering of quantum materials},
  journal = {Annu. Rev. Condens. Matter Phys.},
  volume  = {10},
  pages   = {387--408},
  year    = {2019},
  doi     = {10.1146/annurev-conmatphys-031218-013423}
}

@article{Mori2023,
  author  = {T. Mori},
  title   = {{Floquet} states in open quantum systems},
  journal = {Annu. Rev. Condens. Matter Phys.},
  volume  = {14},
  pages   = {35--56},
  year    = {2023},
  doi     = {10.1146/annurev-conmatphys-040721-015537}
}

@article{Song2020,
  author  = {W.-L. Song and J.-B. You and J. K. Xu and W. L. Yang and J.-H. An},
  title   = {{Floquet} engineering of two weakly coupled superconducting flux qubits},
  journal = {Phys. Rev. Applied},
  volume  = {14},
  pages   = {054049},
  year    = {2020},
  doi     = {10.1103/PhysRevApplied.14.054049}
}

@article{Gallardo2022,
  author  = {S. L. Gallardo and D. Dom{\'i}nguez and M. J. S{\'a}nchez},
  title   = {Dissipative entanglement generation between two qubits parametrically driven and coupled to a resonator},
  journal = {Phys. Rev. A},
  volume  = {105},
  pages   = {052413},
  year    = {2022},
  doi     = {10.1103/PhysRevA.105.052413}
}

@article{Srinivasa2024,
  author  = {V. Srinivasa and J. M. Taylor and J. R. Petta},
  title   = {Cavity-mediated entanglement of parametrically driven spin qubits via sidebands},
  journal = {PRX Quantum},
  volume  = {5},
  pages   = {020339},
  year    = {2024},
  doi     = {10.1103/PRXQuantum.5.020339}
}

@article{Poyatos1996,
  author  = {J. F. Poyatos and J. I. Cirac and P. Zoller},
  title   = {Quantum reservoir engineering with laser cooled trapped ions},
  journal = {Phys. Rev. Lett.},
  volume  = {77},
  pages   = {4728--4731},
  year    = {1996},
  doi     = {10.1103/PhysRevLett.77.4728}
}

@article{Kraus2008,
  author  = {B. Kraus and H. P. B{\"u}chler and S. Diehl and A. Kantian and A. Micheli and P. Zoller},
  title   = {Preparation of entangled states by quantum {Markov} processes},
  journal = {Phys. Rev. A},
  volume  = {78},
  pages   = {042307},
  year    = {2008},
  doi     = {10.1103/PhysRevA.78.042307}
}

@article{Verstraete2009,
  author  = {F. Verstraete and M. M. Wolf and J. I. Cirac},
  title   = {Quantum computation and quantum-state engineering driven by dissipation},
  journal = {Nat. Phys.},
  volume  = {5},
  pages   = {633--636},
  year    = {2009},
  doi     = {10.1038/nphys1342}
}

@article{Krauter2011,
  author  = {H. Krauter and others},
  title   = {Entanglement generated by dissipation and steady state entanglement of two macroscopic objects},
  journal = {Phys. Rev. Lett.},
  volume  = {107},
  pages   = {080503},
  year    = {2011},
  doi     = {10.1103/PhysRevLett.107.080503}
}

@article{Lin2013,
  author  = {Y. Lin and others},
  title   = {Dissipative production of a maximally entangled steady state of two quantum bits},
  journal = {Nature},
  volume  = {504},
  pages   = {415--418},
  year    = {2013},
  doi     = {10.1038/nature12801}
}

@article{Shankar2013,
  author  = {S. Shankar and others},
  title   = {Autonomously stabilized entanglement between two superconducting quantum bits},
  journal = {Nature},
  volume  = {504},
  pages   = {419--422},
  year    = {2013},
  doi     = {10.1038/nature12802}
}

@article{Reiter2013,
  author  = {F. Reiter and L. Tornberg and G. Johansson and A. S. S{\o}rensen},
  title   = {Steady-state entanglement of two superconducting qubits engineered by dissipation},
  journal = {Phys. Rev. A},
  volume  = {88},
  pages   = {032317},
  year    = {2013},
  doi     = {10.1103/PhysRevA.88.032317}
}

@article{Harrington2022,
  author  = {P. M. Harrington and E. J. Mueller and K. W. Murch},
  title   = {Engineered dissipation for quantum information science},
  journal = {Nat. Rev. Phys.},
  volume  = {4},
  pages   = {660--671},
  year    = {2022},
  doi     = {10.1038/s42254-022-00494-8}
}

@article{Beaudoin2011,
  author  = {F. Beaudoin and J. M. Gambetta and A. Blais},
  title   = {Dissipation and ultrastrong coupling in circuit {QED}},
  journal = {Phys. Rev. A},
  volume  = {84},
  pages   = {043832},
  year    = {2011},
  doi     = {10.1103/PhysRevA.84.043832}
}

@article{Ridolfo2012,
  author  = {A. Ridolfo and M. Leib and S. Savasta and M. J. Hartmann},
  title   = {Photon blockade in the ultrastrong coupling regime},
  journal = {Phys. Rev. Lett.},
  volume  = {109},
  pages   = {193602},
  year    = {2012},
  doi     = {10.1103/PhysRevLett.109.193602}
}

@article{Garziano2013,
  author  = {L. Garziano and A. Ridolfo and R. Stassi and Di Stefano, O. and S. Savasta},
  title   = {Switching on and off of ultrastrong light-matter interaction: Photon statistics of quantum vacuum radiation},
  journal = {Phys. Rev. A},
  volume  = {88},
  pages   = {063829},
  year    = {2013},
  doi     = {10.1103/PhysRevA.88.063829}
}

@article{Garziano2017,
  author  = {L. Garziano and A. Ridolfo and De Liberato, S. and S. Savasta},
  title   = {Cavity {QED} in the ultrastrong coupling regime: Photon bunching from the emission of individual dressed qubits},
  journal = {ACS Photonics},
  volume  = {4},
  pages   = {2345--2351},
  year    = {2017},
  doi     = {10.1021/acsphotonics.7b00635}
}

@article{Settineri2018,
  author  = {A. Settineri and V. Macr{\`i} and A. Ridolfo and Di Stefano, O. and A. F. Kockum and F. Nori and S. Savasta},
  title   = {Dissipation and thermal noise in hybrid quantum systems in the ultrastrong-coupling regime},
  journal = {Phys. Rev. A},
  volume  = {98},
  pages   = {053834},
  year    = {2018},
  doi     = {10.1103/PhysRevA.98.053834}
}

@article{Macri2022,
  author  = {V. Macr{\`i} and F. Minganti and A. F. Kockum and A. Ridolfo and S. Savasta and F. Nori},
  title   = {Revealing higher-order light and matter energy exchanges using quantum trajectories in ultrastrong coupling},
  journal = {Phys. Rev. A},
  volume  = {105},
  pages   = {023720},
  year    = {2022},
  doi     = {10.1103/PhysRevA.105.023720}
}

@article{Mercurio2022,
  author  = {A. Mercurio and V. Macr{\`i} and C. Gustin and S. Hughes and S. Savasta and F. Nori},
  title   = {Regimes of cavity {QED} under incoherent excitation: From weak to deep strong coupling},
  journal = {Phys. Rev. Research},
  volume  = {4},
  pages   = {023048},
  year    = {2022},
  doi     = {10.1103/PhysRevResearch.4.023048}
}

@article{FornDiaz2019,
  author  = {P. Forn-D{\'i}az and L. Lamata and E. Rico and J. Kono and E. Solano},
  title   = {Ultrastrong coupling regimes of light-matter interaction},
  journal = {Rev. Mod. Phys.},
  volume  = {91},
  pages   = {025005},
  year    = {2019},
  doi     = {10.1103/RevModPhys.91.025005}
}

@article{Kockum2019,
  author  = {A. F. Kockum and A. Miranowicz and De Liberato, S. and S. Savasta and F. Nori},
  title   = {Ultrastrong coupling between light and matter},
  journal = {Nat. Rev. Phys.},
  volume  = {1},
  pages   = {19--40},
  year    = {2019},
  doi     = {10.1038/s42254-018-0006-2}
}

@article{Kohler1997,
  author  = {S. Kohler and T. Dittrich and P. H{\"a}nggi},
  title   = {{Floquet}-{Markovian} description of the parametrically driven, dissipative harmonic oscillator},
  journal = {Phys. Rev. E},
  volume  = {55},
  pages   = {300--313},
  year    = {1997},
  doi     = {10.1103/PhysRevE.55.300}
}

@article{Gasparinetti2013,
  author  = {S. Gasparinetti and P. Solinas and S. Pugnetti and R. Fazio and J. P. Pekola},
  title   = {Environment-governed dynamics in driven quantum systems},
  journal = {Phys. Rev. Lett.},
  volume  = {110},
  pages   = {150403},
  year    = {2013},
  doi     = {10.1103/PhysRevLett.110.150403}
}

@article{Hone2009,
  author  = {D. W. Hone and R. Ketzmerick and W. Kohn},
  title   = {Statistical mechanics of {Floquet} systems: The pervasive problem of near degeneracies},
  journal = {Phys. Rev. E},
  volume  = {79},
  pages   = {051129},
  year    = {2009},
  doi     = {10.1103/PhysRevE.79.051129}
}

@misc{Akbari2026,
  author        = {K. Akbari and F. Nori and S. Hughes},
  title         = {{Floquet} quasienergy-resolved dissipation, dynamics, and spectroscopy in ultrastrong cavity-{QED}},
  year          = {2026},
  eprint        = {2606.31108},
  archiveprefix = {arXiv},
  primaryclass  = {quant-ph},
  doi           = {10.48550/arXiv.2606.31108}
}

@article{Gorini1976,
  author  = {V. Gorini and A. Kossakowski and E. C. G. Sudarshan},
  title   = {Completely positive dynamical semigroups of {$N$}-level systems},
  journal = {J. Math. Phys.},
  volume  = {17},
  pages   = {821--825},
  year    = {1976},
  doi     = {10.1063/1.522979}
}

@article{Lindblad1976,
  author  = {G. Lindblad},
  title   = {On the generators of quantum dynamical semigroups},
  journal = {Commun. Math. Phys.},
  volume  = {48},
  pages   = {119--130},
  year    = {1976},
  doi     = {10.1007/BF01608499}
}

@article{Farina2019,
  author  = {D. Farina and V. Giovannetti},
  title   = {Open-quantum-system dynamics: Recovering positivity of the {Redfield} equation via the partial secular approximation},
  journal = {Phys. Rev. A},
  volume  = {100},
  pages   = {012107},
  year    = {2019},
  doi     = {10.1103/PhysRevA.100.012107}
}

@article{NathanRudner2020,
  author  = {F. Nathan and M. S. Rudner},
  title   = {Universal {Lindblad} equation for open quantum systems},
  journal = {Phys. Rev. B},
  volume  = {102},
  pages   = {115109},
  year    = {2020},
  doi     = {10.1103/PhysRevB.102.115109},
  note    = {Erratum: Phys. Rev. B 104, 119901 (2021)}
}

@article{Mozgunov2020,
  author  = {E. Mozgunov and D. A. Lidar},
  title   = {Completely positive master equation for arbitrary driving and small level spacing},
  journal = {Quantum},
  volume  = {4},
  pages   = {227},
  year    = {2020},
  doi     = {10.22331/q-2020-02-06-227}
}

@article{Trushechkin2021,
  author  = {A. Trushechkin},
  title   = {Unified {Gorini}-{Kossakowski}-{Lindblad}-{Sudarshan} quantum master equation beyond the secular approximation},
  journal = {Phys. Rev. A},
  volume  = {103},
  pages   = {062226},
  year    = {2021},
  doi     = {10.1103/PhysRevA.103.062226}
}

@article{Davies1974,
  author  = {E. B. Davies},
  title   = {Markovian master equations},
  journal = {Commun. Math. Phys.},
  volume  = {39},
  pages   = {91--110},
  year    = {1974},
  doi     = {10.1007/BF01608389}
}

@article{GardinerCollett1985,
  author  = {C. W. Gardiner and M. J. Collett},
  title   = {Input and output in damped quantum systems: Quantum stochastic differential equations and the master equation},
  journal = {Phys. Rev. A},
  volume  = {31},
  pages   = {3761--3774},
  year    = {1985},
  doi     = {10.1103/PhysRevA.31.3761}
}

@article{CiutiCarusotto2006,
  author  = {C. Ciuti and I. Carusotto},
  title   = {Input-output theory of cavities in the ultrastrong coupling regime: The case of time-independent cavity parameters},
  journal = {Phys. Rev. A},
  volume  = {74},
  pages   = {033811},
  year    = {2006},
  doi     = {10.1103/PhysRevA.74.033811}
}

@article{Settineri2021,
  author  = {A. Settineri and Di Stefano, O. and D. Zueco and S. Hughes and S. Savasta and F. Nori},
  title   = {Gauge freedom, quantum measurements, and time-dependent interactions in cavity {QED}},
  journal = {Phys. Rev. Research},
  volume  = {3},
  pages   = {023079},
  year    = {2021},
  doi     = {10.1103/PhysRevResearch.3.023079}
}

@article{Macieszczak2016,
  author  = {K. Macieszczak and M. Gu{\c{t}}{\u{a}} and I. Lesanovsky and J. P. Garrahan},
  title   = {Towards a theory of metastability in open quantum dynamics},
  journal = {Phys. Rev. Lett.},
  volume  = {116},
  pages   = {240404},
  year    = {2016},
  doi     = {10.1103/PhysRevLett.116.240404}
}

@article{Minganti2018,
  author  = {F. Minganti and A. Biella and N. Bartolo and C. Ciuti},
  title   = {Spectral theory of {Liouvillians} for dissipative phase transitions},
  journal = {Phys. Rev. A},
  volume  = {98},
  pages   = {042118},
  year    = {2018},
  doi     = {10.1103/PhysRevA.98.042118}
}

@article{Wootters1998,
  author  = {W. K. Wootters},
  title   = {Entanglement of formation of an arbitrary state of two qubits},
  journal = {Phys. Rev. Lett.},
  volume  = {80},
  pages   = {2245--2248},
  year    = {1998},
  doi     = {10.1103/PhysRevLett.80.2245}
}

@article{Bluemel1991,
  author  = {R. Bl{\"u}mel and A. Buchleitner and R. Graham and L. Sirko and U. Smilansky and H. Walther},
  title   = {Dynamical localization in the microwave interaction of {Rydberg} atoms: The influence of noise},
  journal = {Phys. Rev. A},
  volume  = {44},
  pages   = {4521--4540},
  year    = {1991},
  doi     = {10.1103/PhysRevA.44.4521}
}

@misc{Mercurio2026DPT,
  author        = {A. Mercurio and V. Macr{\`i} and F. Ferrari and L. Fioroni and V. Savona},
  title         = {{Floquet} dissipative phase transitions},
  year          = {2026},
  eprint        = {2603.13030},
  archiveprefix = {arXiv},
  primaryclass  = {quant-ph},
  doi           = {10.48550/arXiv.2603.13030}
}

@article{Clerk2010,
  author  = {Clerk, A. A. and Devoret, M. H. and Girvin, S. M. and Marquardt, Florian and Schoelkopf, R. J.},
  title   = {Introduction to quantum noise, measurement, and amplification},
  journal = {Rev. Mod. Phys.},
  volume  = {82},
  pages   = {1155--1208},
  year    = {2010},
  doi     = {10.1103/RevModPhys.82.1155}
}
\end{document}